\documentclass[12pt]{article}
\usepackage{authblk}
\usepackage{natbib}
\usepackage[right=2.5cm,left=2.5cm,margin=3cm]{geometry}

\usepackage{amssymb,amsmath,bm, mathtools}
\usepackage[utf8x]{inputenc}
\usepackage[T1]{fontenc}
\usepackage{siunitx}
\usepackage[table]{xcolor}
\usepackage{float}
\usepackage{graphicx}
\usepackage[unicode=true]{hyperref}
\usepackage{threeparttable}
\usepackage{caption}
\usepackage{subcaption}

\hypersetup{breaklinks=true,
            colorlinks=false,
            pdfborder={0 0 0}}

\renewcommand{\Re}{\mathbb{R}}

\newcommand{\Co}{\mathbb{C}}
\newcommand{\mat}[1]{\boldsymbol{#1}}
\renewcommand{\vec}[1]{\boldsymbol{#1}}

\newcommand{\tr}{^{\mathrm{T}}}
\newcommand{\trH}{^{\mathrm{H}}}

\newcommand{\tamanhoR}[1]{\in\Re^{#1}}
\newcommand{\tamanhoC}[1]{\in\Co^{#1}}

\newcommand{\media}[1]{\bar{\vec{#1}}}

\renewcommand{\eqref}[1]{(\ref{#1})}
\newcommand{\figref}[1]{Figure~\ref{#1}}
\newcommand{\subfigref}[2]{Figure~\ref{#1}#2 }
\newcommand{\tabref}[1]{Table~\ref{#1}}
\newcommand{\secref}[1]{Section~\ref{#1}}

\newcommand{\sumS}{\sum_{s=1}^{N_S}}
\newcommand{\sumI}{\sum_{i=1}^{N_I}}

\newcommand{\image}[5][h]{
\begin{figure}[#1]
  \center
  \includegraphics[width=#5]{./images/#2}
  \caption{#3}\label{#4}
\end{figure}
}

\setcitestyle{aysep={},citesep={,}}

\title{Anatomical atlas of the upper part of the human head for electroencephalography and bioimpedance applications}

{\small
\author[1,2,*]{Fernando S Moura}

\author[1]{ Roberto G Beraldo}

\author[1]{ Leonardo A Ferreira}

\author[2]{ Samuli Siltanen}
\affil[1]{{\small Engineering, modelling and Applied Social Sciences Center, Federal University of ABC\\ S\~ao Bernardo do Campo, S\~ao Paulo, Brazil}}
\affil[2]{{\small Department of Mathematics and Statistics, University of Helsinki, Helsinki, Finland}}
\affil[*]{{\small Corresponding author. Email: fernando.moura@ufabc.edu.br}}
}
\date{}
\begin{document}
\maketitle

\begin{abstract}
Volume conductor problems in cerebral electrophysiology and bioimpedance do not have analytical solutions for nontrivial geometries and require a 3D model of the head and its electrical properties for solving the associated PDEs numerically.
Ideally, the model should be made with patient-specific information. In clinical practice, this is not always the case and an average head model is often used. Also, the electrical properties of the tissues might not be completely known due to natural variability.
The objective of this work is to develop a 4D (3D+T) statistical anatomical atlas of the electrical properties of the upper part of the human head for cerebral electrophysiology and bioimpedance applications.
The atlas is an important tool for \textit{in silico} studies on cerebral circulation and electrophysiology that require statistically consistent data, e.g., machine learning, sensitivity analyses, and as a benchmark to test inverse problem solvers.
The atlas was constructed based on MRI images of human individuals and comprises the electrical properties of the main internal structures and can be adjusted for specific electrical frequencies.
The proposed atlas also comprises a time-varying model of arterial brain circulation, based on the solution of the Navier-Stokes equation in the main arteries and their vascular territories.
The atlas was successfully used to simulate electrical impedance tomography measurements indicating the necessity of signal-to-noise between 100 and \SI{125}{\decibel} to identify vascular changes due to the cardiac cycle, corroborating previous studies.
\end{abstract}



\section{Introduction}

Electrophysiology is the branch of physiology that investigates the electrical properties of biological tissues. The analysis is based on electrical measurements, voltages, or electric currents, generated by the tissue or in response to external electric stimuli.

One special group is clinical neurophysiology, where the bioelectrical activity is recorded to assess central and peripheral nervous systems. Electroencephalography (EEG) is an important monitoring and diagnostic method in this speciality to record brain electrical activity that can be used to diagnose thalamocortical rhythms, such as assessing seizure disorders, epilepsy, sleep disorders, coma, schizophrenia, Parkinson disease, and brain death \citep{Michel2019,jatoi2017brain}.

Electroencephalography measures voltage variations using multiple electrodes typically placed along the scalp of the patient. The measured voltages are the result of ionic currents inside the brain, therefore they are caused by spontaneous electrical activity.

In the case of epilepsy diagnostics, EEG is employed to determine the type, location, and extension of the lesion causing seizures. This is a challenging task since it depends on measurements taken on the surface of the scalp to infer the internal source of the disorder, a process called source reconstruction \citep{Hallez2007, Grech2008}.

Bioimpedance analysis is another group of methods used to assess the electrical properties of biological tissues. Measurements are made in response to external stimuli, such as measuring voltages caused by external current sources attached to the surface of the body or vice-versa.

Electrical impedance tomography (EIT) is a medical image technique in which electrical measurements on the surface of the body are used to create an image of conductivity distribution (or admittivity, the complex-valued equivalent) within the body. The image can then be associated with the physiological conditions of the patient. EIT has been used successfully in many areas, such as in lung applications to monitor acute respiratory distress syndrome, obstructive lung diseases or perioperative patients \citep{MARTINS2019442}, monitoring mechanical ventilation, heart activity \citep{Frerichs2016}, cardiac function and detecting cancerous tissues \citep{Adler2019}. It is also being investigated to monitor brain activity and to distinguish between ischemic and hemorrhagic stroke \citep{Adler2019}.

Many electrophysiology applications require solving a nonlinear ill-posed inverse problem associated with the volume conductor. Reliable and stable solutions depend on prior information about the geometry and electrical properties of the tissues and knowledge about measurement uncertainty. In the case of source reconstruction, prior information about the electrical properties of the tissues is required for composing the forward volume conduction model needed in the process. Small errors in the electrical properties inside the head can obfuscate the effects of deep brain activity. In the case of EIT, anatomical and electrical prior information is also required to restrict the solution search space.

In addition, the brain is not a static structure. The flow of blood varies periodically during each cardiac cycle \citep{Wagshul2011}. Intracranial pulsatility has been evaluated using magnetic resonance imaging \citep{Vikner2019, Holmgren2019}, or transcranial Doppler sonography \citep{Kneihsl2020}, and tissue pulsatility imaging \citep{Kucewicz2008, Desmidt2018}.

Blood flow centre-line velocity and artery radius influence the electrical impedance of blood in that artery \citep{Gaw2008,Shen2016,Shen2018}. This makes it promising to use electrical conductivity measurements, such as in EIT or impedance cardiography, in haemodynamic monitoring using surface electrodes on the skin. In fact, there have been several works recently aiming to monitor blood flow and/or pressure waveforms using them, e.g., common carotid arteries \citep{Zhang2020}, aortic artery \citep{Badeli2020}, pulmonary artery \citep{Braun_2018}, radial artery \citep{Pesti2019}, and cerebral arteries \citep{rgb}. There are works about impedance cardiography to determine stroke volume \citep{Impedancecardiography}, electrical bioimpedance sensing to determine the central aortic pressure (CAP) curves \citep{Min2019}, and pulmonary artery pressure estimation using EIT \citep{Proena2020}. There is an increasing interest in brain monitoring using electrical measurements, such as to monitor ventricular volume \citep{CastelarWembers:759058}, rheoencephalography to assess cerebral blood flow \citep{Bodo2018, Meghdadi2019}, brain perfusion of rats \citep{Dowrick2016, Song2018}, and stroke identification \citep{Goren2018,agnelli2020classification,candiani2020neural,candiani2019computational}.

Volume conductor problems in electrophysiology and bioimpedance do not have analytical solutions for nontrivial geometries and rely on numerical methods, e.g., finite element method (FEM) to discretize the head in small elements and solve the associated PDEs. Ideally, the FEM model should be built with patient-specific information, taken from MRI or CT scans to capture precisely the geometry of the head, its internal structures, and electrode positions. Unfortunately, in clinical practice, this is not the case. Often, an oversimplified geometry is employed for all patients due to the lack of computational tools and time.

The effects of mismodelling have been investigated before. EEG source localization errors increase substantially when individual-specific head models are not at disposal \citep{AkalinAcar2013}. The authors also show that errors in the conductivity of the skull cause large estimate errors. The latter is especially challenging because the electrical properties of the skull are highly heterogeneous and have large variability inter-individual. Cerebrospinal fluid (CSF) has a big impact on the results due to its high conductivity that forms a conductive layer surrounding the brain effectively shielding the interior \citep{Vorwerk2014,Cho2015}. Also, the authors show that distinguishing white and grey matters also impact the head volume conductor model.

The objective of this work is to develop a statistical anatomical atlas of electrical properties of the upper part of the human head for electrophysiology and bioimpedance applications. The atlas is constructed based on MRI images of human individuals and comprises the electrical properties of the main structures for electrophysiology. The proposed atlas also comprises a time-varying model of the brain circulation, based on the solution of the Navier-Stokes equation for  blood flow in the main arteries \citep{Melis2017,openBF.jl-2018}. The atlas can be used to generate synthetic data statistically consistent with the population to compose learning sets for machine learning methods, for sensitivity analyses, and as a benchmark to test algorithms. The atlas can also be used as statistical prior information for inverse problems in electrophysiology. Anatomy-based priors are found in the literature, such as for thorax applications \citep{MARTINS2019442,kaipio2005statistical}.

\section{Anatomical Atlas Description and Construction}

The anatomical atlas is composed of a static component $A_s$ with the electrical properties of the main tissues found in the upper part of the human head and a dynamic component $A_d(t)$ to account for blood perfusion dynamics in the human head. The two components of the atlas are considered Gaussian and independent, therefore the final statistics of the atlas $A$ is  composed by 
\begin{align}
    A_s &\sim N(\media x_s, \mat\Gamma_s)\\
    A_d(t) &\sim N(\media x_d(t), \mat\Gamma_d(t))\\
    A(t) &\sim N(\media x_s+\media x_d(t), \mat \Gamma_a+\mat\Gamma_d(t)).
\end{align}

The two components of the atlas are presented in details in the following subsections.

\subsection{Static component} \label{sec:atlasStatic}

The static component of the atlas distinguishes five main compartments of importance for electrophysiology of the human head: grey matter (GM), white matter (WM), cerebrospinal fluid (CSF), bones (BO) and other soft tissues (OT).  \figref{fig:staticAtlasDiagram} depicts the general procedure to calculate the static component of the anatomical atlas.

\image[ht]{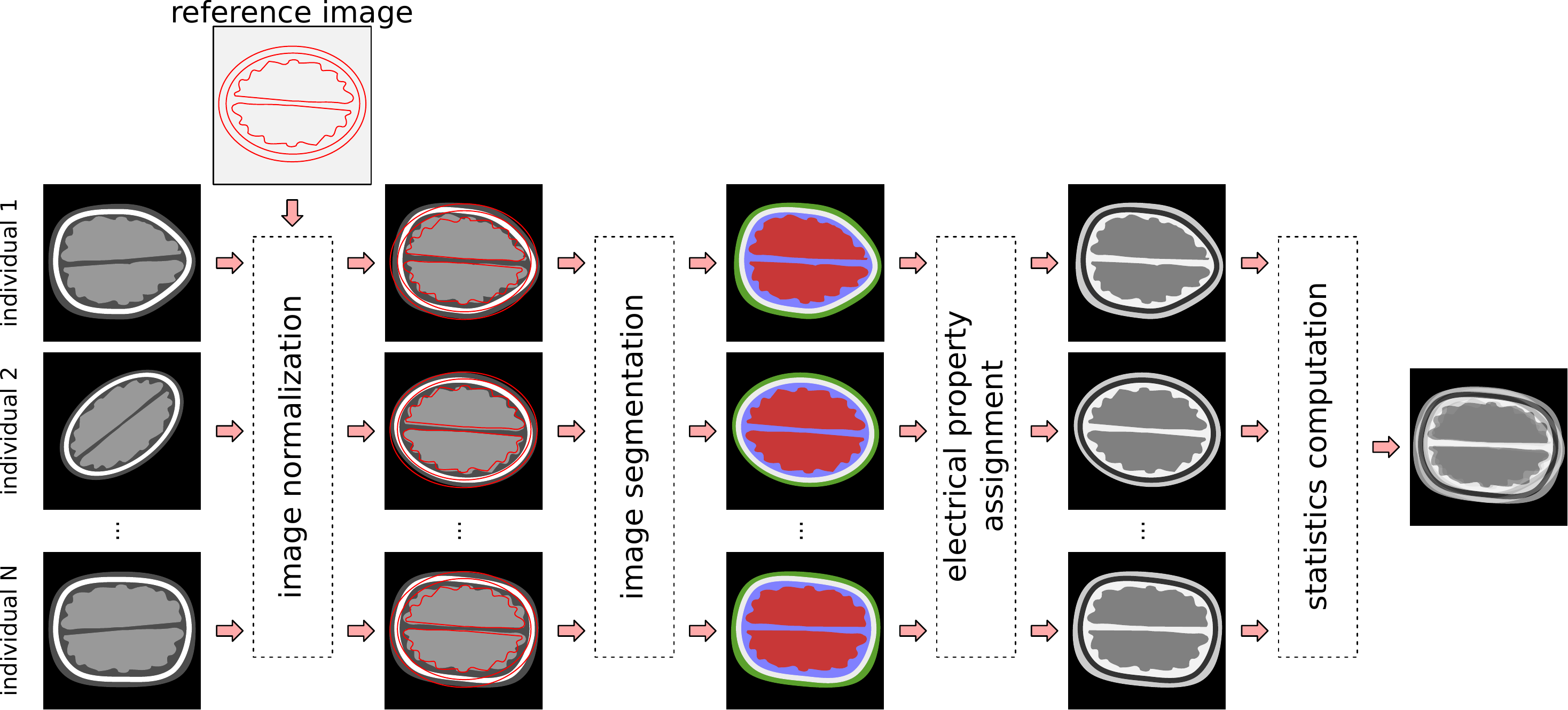}{Main steps necessary to compute the static component of the anatomical atlas. Each image is normalized to a reference image, segmented into the main compartments. Each segment is assigned with the corresponding electrical property and, finally, the statistics of the atlas can be estimated.}{fig:staticAtlasDiagram}{\textwidth}

Fifty 3D Magnetic Resonance (MR) images of healthy human individuals, made available by the CASILab at the University of North Carolina at Chapel Hill were used \citep{mha}. MR images are of type T1, obtained in a three-tesla equipment using the Fast Low Angle Shot (FLASH) sequence with a resolution of \SI[product-units = single]{1 x 1 x 1}{mm}. An equal number of male and female individuals were used, with an average age of $40\pm15$ years old.

The Symmetric image Normalization (SyN) method was applied to the images to diminish differences due to misalignment, aspect ratio, and sizes between the heads \citep{AVANTS2008}. For this purpose, the Advanced Normalization Tools (ANTs) was used, under the Neuroimaging in Python Pipelines and Interfaces (Nipype) framework \citep{nipype}. Detailed information regarding the normalization can be found in \cite{AVANTS2008,avants2009ants}.

Each of the 50 images was transformed aiming to maximize its similarity with a reference image. The reference image is the MNI ICBM 152\footnote{Available at http://nist.mni.mcgill.ca/?p=858}, a nonlinear symmetric atlas with  \SI[product-units = single]{1 x 1 x 1}{mm} resolution \citep{Grabner_2006,fonov2009unbiased}. The reference image is presented in \figref{fig:refImgICB152}. Using an average head geometry as reference avoids having to choose one of the images in the dataset as reference, eliminating the possibility of choosing as a reference an individual with any abnormal geometric feature. Each transformation is performed in two stages, first a rigid transformation to roughly align the geometries, followed by an affine transformation.

\image[H]{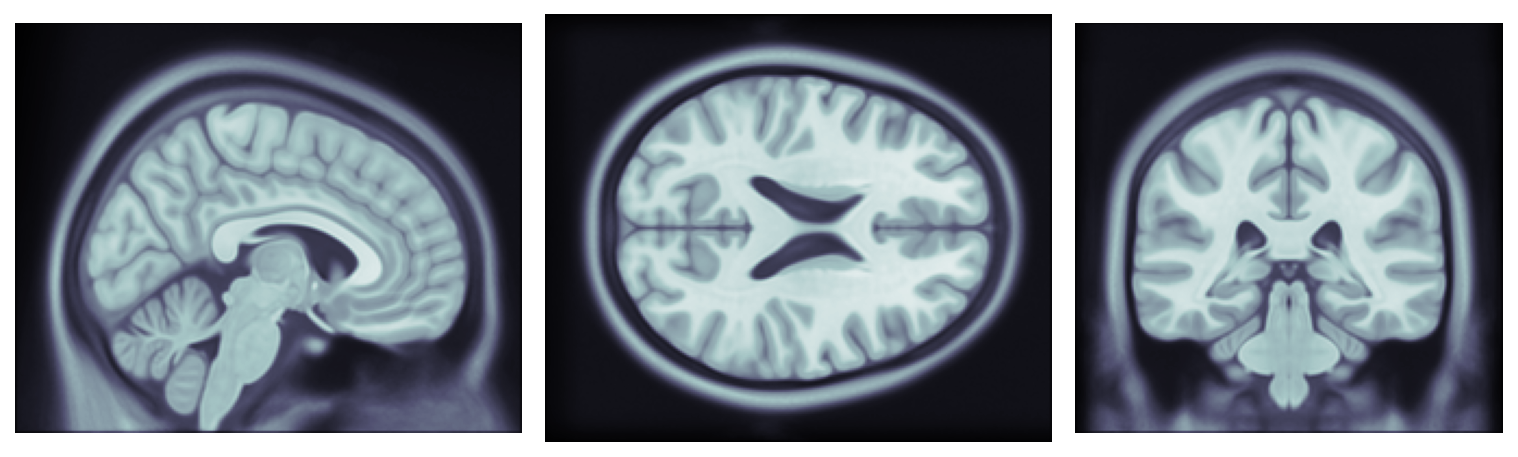}{Slices of the reference image MNI ICBM 152.}{fig:refImgICB152}{0.7\textwidth}

The normalization process assumes the heads have similar shapes and proportions. Therefore, the atlas represents a head with size and proportions similar to the reference head used for the normalization. If necessary, the resulting atlas can be transformed to accommodate other geometries, for example when the geometry of the head of the patient is available or if an average head model is preferred. This procedure will be described in \secref{sec:exampleEIT}

After the spatial normalization, the images were segmented into six classes: background, WM, GM, CSF, BO, and OT. The Statistical Parametric Mapping (SPM) was used to segment the images \citep{SPM}. The method is composed of three distinct components that are combined and optimized circularly: (i) modelling of the intensities of the images using a Gaussian mixture model; (ii) normalization of tissue probability maps of the five parts of the head with the images; and (iii) a bias field correction. Further details about the implementation can be found in \cite{ashburner2005unified}. At the end of this phase, each voxel of the images is assigned to the label with the highest probability.

Three additional steps were also performed to improve segmentation. (i) Any segmentation holes inside the head were filled with the nearest tissue in the image. This procedure was applied to all 2D slices in the three anatomical planes of each image, (ii) Small segmentation artefacts outside the human head were removed by isolating the largest connected group in the image using a six-connected neighbourhood strategy. (iii) Four iterations of morphological opening operation to the binary mask of the head to smooth the external surface of the head.

\subsubsection{Electrical properties of the segments}

Each voxel of the segmented images was assigned to the electrical property of the corresponding tissue before computing the statistics of the atlas. Tissues were modelled as isotropic, even though it is known that some tissues are anisotropic. The electrical properties depend on the type, physiological conditions and frequency in consideration \citep{gabriel1,gabriel2}.

Given the angular frequency of the electrical signal $\omega=2\pi f$ , the complex relative permittivity $\hat \epsilon(\omega)$ of a tissue can be modelled as the sum of four Cole-Cole dispersion terms \citep{gabriel3}
\begin{align}
    \hat \epsilon(\omega)=\epsilon_{\infty} +\sum_{n=1}^4\frac{\Delta \epsilon_n}{1+(j\omega\tau_n)^{1-\alpha_n}} + \frac{\sigma_0}{j\omega\epsilon_0},\label{eq:prop}
\end{align}
where $\epsilon_0$ is the permittivity of free space and all the other parameters depend on the tissue \citep{gabriel3,Andreuccetti}. The conductivity $\sigma$ and permittivity $\epsilon$ of the tissue can be obtained from $\hat \epsilon(\omega)$
\begin{align}
    \sigma(\omega)&=-\omega \epsilon_0 \text{Im}(\hat \epsilon(\omega)) \\
    \epsilon(\omega) &= \epsilon_0 \text{Re}(\hat \epsilon(\omega)).
\end{align}

Biological tissues are naturally inhomogeneous, due to their complex macroscopic and microscopic structure, function and physiological condition. To account for this, the uncertainty level of the estimates from the above equations was set to $\pm 20\%$ following reported results in \cite{gabriel2}.

The Electrical properties of BO were modelled as the average between cortical and cancellous bones, while OT was modelled as muscle tissue.

\subsubsection{Atlas statistics computation}\label{sec:statistics}

Let $\vec u \tamanhoR{N_V}$ be a vector representing a 3D image of the human head after normalization and segmentation, where $N_V$ is the number of voxels, excluding those representing the background around the head. Let the image be segmented into $N_T$ nonintersecting regions (tissues), each one with associated characteristic funcion $\vec{\chi}_t \tamanhoR{N_V}$, for $t=1,2,\cdots,N_T$. Also, let $p_t$ be the electrical property of each tissue under consideration. The property can be real (e.g., resistivity or conductivity) or complex valued (e.g., impeditivity or admitivity). We can write the 3D image of this property $\vec{x}\tamanhoC{N_T}$ (or real with the same dimension) as
\begin{align}
\vec x  = \sum_{t=1}^{N_T} p_t\vec{\chi}_t = \mat X\vec{p},\label{eq:img_rho}
\end{align}
where $\vec{p}\tamanhoC{N_T}$ is a vector composed by the electrical properties of the tissues and $\mat X\tamanhoR{N_V\times N_T}$ is a matrix where each column is a characteristic function.

Assume images from $N_I$ individuals are used to build the atlas. Formally this number should be very large to represent the statistics of the population. In practice, this number is limited by the size of the dataset. To reduce this limitation each individual will be considered $N_S$ times, each time with a different value for $\vec{p}$, following the statistics of the tissues. This implies that the $N_I$ individuals represent the general shape of the head of the population while allowing the electrical properties of the tissues to be more diverse. In addition, we assume the same number of samples $N_S$ per individual, making them equally probable.

Using these hypotheses the average and covariance of the population can be estimated efficiently. For the covariance matrix $\mat \Gamma$ in special, the formulation allows the computation in factorized form $\mat \Gamma = \mat K \mat K\tr$, reducing storage requirements and simplifying algorithms that depend on factorizations of $\mat \Gamma$.

Samples of the $i$-th individual can be composed by sampling the properties of the tissues $\vec p_s$ and applying \eqref{eq:img_rho}
\begin{align}
\vec x_{s,i}  = \mat X_i\vec{p}_s, \quad \text{for}\quad s=1,2,\cdots,N_S,\label{eq:sampleX}
\end{align}
where the samples $\vec{p}_s$ can be generated from data fitted models or measurements. In this work, the electrical properties of the tissues are considered Gaussian with average resulting from the model \eqref{eq:prop} and standard deviation of 20\% of the average, following reported results \citep{gabriel2}.

Let $\vec x_{s,i}$ represent a sample of the $i$-th individual. The average over all individuals can be estimated with
\begin{equation}
    \media{x} = \frac{1}{N_S N_I}\sumI  \sumS \vec x_{s,i}=
                    \frac{1}{N_I}\sumI \mat X_i\frac{1}{N_S}\sumS\vec{p}_s=
                    \frac{1}{N_I} \sumI \mat X_i\media{p},\label{eq:atlas_avg}
\end{equation}
where, again, $N_S$ is fixed and represent the number of samples with the same head geometry $\mat X_i$ and $\media{p}$ is the average electrical properties of the tissues.

The covariance matrix can be computed using the usual sample estimator
\begin{equation}
    \mat\Gamma = \frac{1}{N_I N_S-1} \sumI  \sumS (\vec x_{s,i} - \media{x} ) (\vec x_{s,i} - \media{x})\trH,
\end{equation}
where $\mat\Gamma\tamanhoR{N_V\times N_V}$ and $\mat M\trH$ denotes conjugate transpose of $\mat M$. For real valued $\vec p$, the conjugate transpose is the transpose $\mat M\trH=\mat M\tr$

Adding $(\media{x}_i-\media{x}_i)$ to both terms between parenthesis and rearranging the terms,
\begin{equation}
    \mat\Gamma = \frac{1}{N_I N_S-1} \sumI  \sumS [(\vec x_{s,i}-\media{x}_i) + (\media{x}_i - \media{x}) ] [(\vec x_{s,i}-\media{x}_i)\trH +(\media{x}_i- \media{x})\trH],
\end{equation}
where $\media{x}_i=\sumS \vec x_{s,i} /N_S$ is the average of the $i$-th individual. Proceeding with the products, 
\begin{align}
    \mat\Gamma &= \frac{1}{N_I N_S-1} \sumI (\alpha_1+\alpha_2+\alpha_3 + \alpha_4)\\
    \alpha_1&=\sumS(\vec x_{s,i}-\media{x}_i) (\vec x_{s,i}-\media{x}_i)\trH =
              (N_S-1) \mat\Gamma_i\\
    \alpha_2&=\sumS(\vec x_{s,i}-\media{x}_i)(\media{x}_i- \media{x})\trH=
              \left[\sumS(\vec x_{s,i}-\media{x}_i)\right](\media{x}_i- \media{x})\trH=0\\
    \alpha_3&=\sumS(\media{x}_i - \media{x})(\vec x_{s,i}-\media{x}_i)\trH=
              (\media{x}_i - \media{x})\left[\sumS(\vec x_{s,i}-\media{x}_i)\trH\right]=0\\
    \alpha_4&=\sumS(\media{x}_i - \media{x})(\media{x}_i- \media{x})\trH=
              N_S(\media{x}_i - \media{x})(\media{x}_i- \media{x})\trH.
\end{align}

Finally, taking the limit $N_S\rightarrow \infty$,
\begin{align}
    \mat\Gamma &=\frac{1}{N_I} \sumI\mat\Gamma_i + \frac{1}{N_I} \sumI (\media{x}_i - \media{x})(\media{x}_i- \media{x})\trH.\label{eq:atlas_cov1}
\end{align}
Defining $\Delta \media{x}_i=\media{x}_i- \media{x}$ and observing the linear relation \eqref{eq:img_rho} we can rewrite this last expression as
\begin{align}
    \mat\Gamma &=\frac{1}{N_I} \sumI[\mat X_i\mat \Gamma_p\mat X_i\trH + \Delta \media{x}_i\Delta \media{x}_i\trH]=
                 \sumI\mat W_i\mat W_i\trH\\
    \mat W_i&=\frac{1}{\sqrt{N_I}}\begin{bmatrix}\mat X_i\sqrt{\mat \Gamma_p}  &\Delta \media{x}_i\end{bmatrix},\label{eqref:Wi}
\end{align}
where $\mat W_i\tamanhoC{N_V\times (N_T+1)}$ and $\mat \Gamma_p\tamanhoC{N_T\times N_T}$ is the covariance matrix of the electrical properties of the tissues. Furthermore, the expression can be simplified to
\begin{align}
    \mat\Gamma &=\begin{bmatrix}\mat W_1 & \cdots & \mat W_I\end{bmatrix}
    \begin{bmatrix}\mat W_1\trH \\ \vdots \\ \mat W_I\trH\end{bmatrix}=\mat K\mat K\trH,\label{eq:atlas_cov}
\end{align}
where $\mat K\tamanhoC{N_V\times [(N_T+1)N_I]}$. Note that both the average \eqref{eq:atlas_avg} and the covariance \eqref{eq:atlas_cov} estimates do not require explicit sampling procedure presented in \eqref{eq:sampleX}.

In case of complex-valued $\vec p$, the pseudo-covariance $\mat{\tilde\Gamma}\tamanhoC{N_V\times N_V}$ can be computed in a similar way
\begin{align}
    \mat{\tilde\Gamma} &=\mat K\mat K\tr\label{eq:atlas_Pcov}.
\end{align}

The atlas in this work is assumed Gaussian, therefore $\media{x}$ and $\mat\Gamma$ (and $\mat{\tilde\Gamma}$ for complex-valued $\vec p$) completely specify its probability density function.

\subsection{Dynamic component}\label{sec:segmentationD}

Flow in the cranial cavity is pulsatile, following the cardiac cycle. Arterial blood flows in waves, forcing part of the venous blood and CSF to move, following the Monro–Kellie hypothesis \citep{Wagshul2011}. Venous blood is drained to the jugular veins via cerebral sinuses, while CSF moves in the subarachnoid space and leaves/returns to the cavity via the foramen magnum to balance intracranial pressure waves during the cardiac cycle \citep{Greitz1992,Sakka2011}. Recently, pulsatility was also observed in small cortical veins \citep{DRIVER_2020}, and studies in rats show that the flow in microvessels is quasi-steady laminar flow, following Hagen–Poiseuille law expected in low Reynolds and Womersley numbers \citep{JUNJI_2006}.

Arterial blood enters the cranial cavity through its base via two pairs of arteries, the (right/left) vertebral and internal carotid arteries. After entering, these arteries form the circle of Willis, a circulatory anastomosis responsible for providing backup routes for cerebral blood supply \citep{Bradac2017,circulation1}. From the circle of Willis four main pairs of arteries branch out, the (right/left) anterior, middle, posterior cerebral, and superior cerebellar arteries.

The dynamic component of the atlas comprises the circulation in the main cerebral arteries. The procedure follows the same main steps presented in \secref{sec:atlasStatic} with a few modifications. (i) Magnetic Resonance Angiography (MRA) images of 109 healthy human individuals were used \citep{mha}. The images were obtained in a three tesla equipment with a resolution of \SI[product-units = single]{0.5 x 0.5 x 0.8}{mm}. An equal number of male and female individuals were used, with an average age of $43\pm14$ years old. (ii) Segmentation was performed first by applying a total variation filter to the images followed by a threshold segmentation. Only the lumen of the vessels with contrast agent were segmented. (iii) Electrical property assignment follows the procedure described in the following subsection.

\subsubsection{Electrical properties of the segments}

The influence of blood flow centre-line velocity and vessel radius over the electrical impedance of blood is modelled and included in the atlas \citep{Gaw2008,Shen2016,Shen2018}.

We simulated blood flow in the main arteries of the brain using the openBF solver \citep{Melis2017,openBF.jl-2018}, a 1D blood flow solver based on monotonic upstream-centered scheme for conservation laws (MUSCL) finite-volume numerical scheme. The solver assumes the blood is an incompressible Newtonian fluid, flowing through narrow and long circular vessels with linear compliant walls. The Navier-Stokes equations are reduced to 1D by imposing axisymmetry, linearized and solved for pulsatile flows using the finite difference method. Detailed description can be found in \cite{MelisThesis}.

Brain circulation simulation encompasses the superior aortic system, from the ascending aorta to the main arteries providing blood to the brain. The arteries considered in the simulation can be seen in \figref{fig:1Dmodel_01}.

\image[H]{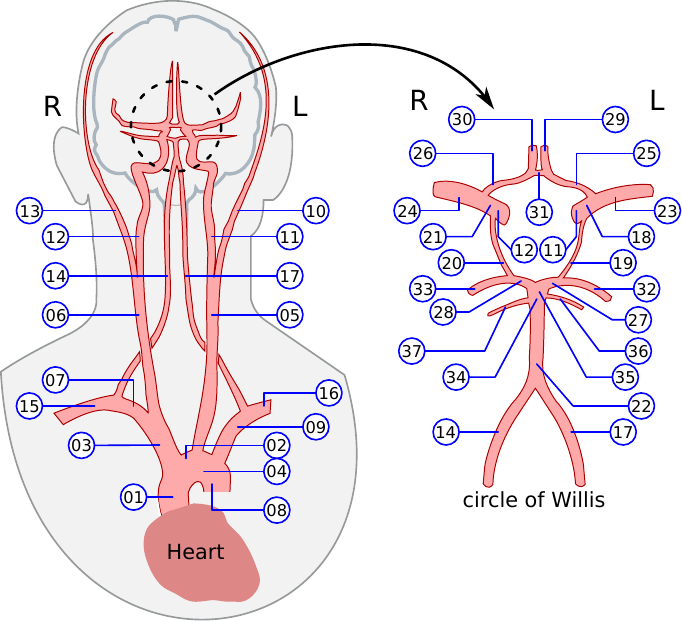}{Superior aortic system considered in the simulations. The names of the vessels are presented in \tabref{tab:vessels}.}{fig:1Dmodel_01}{0.6\textwidth}

Blood was assumed Newtonian with density $\rho=\SI{1050}{\kilo\gram\per\metre\cubed}$ and dynamic viscosity $\mu=\SI{4.5e-3}{\pascal\second}$. The geometry and mechanical properties of the vessels are presented in \tabref{tab:vessels}, based on \cite{Alastruey2007} and complemented with data collected by \cite{Dodo2020,Fomkina2016,Schmitter2013}. The terminal vessels were coupled with 3-element Windkessel models to mimic the perfusion of downstream vessels and avoid numerical oscillations. Heart flow output in one cardiac cycle was set to
\begin{align}
    Q(t)=\left\{\begin{array}{ll}Q_M\sin(\frac{\pi t}{\tau})& t<\tau\\0& \text{otherwise}\end{array}\right.,
\end{align}
where $Q_M=\SI{485}{\milli\litre\per\second}$ is peak flow rate, $\tau=\SI{0.3}{\second}$ and the cardiac cycle period is \SI{1}{\second}, following \cite{Alastruey2007}.

\begin{table}[H]
\rowcolors{2}{gray!15}{white}
{\scriptsize
\centering
 \begin{threeparttable}[t]
\caption{Geometrical and mechanical properties of the arteries. Based on \cite{Alastruey2007} and complemented with data collected by \cite{Dodo2020,Fomkina2016,Schmitter2013}. $\ell:$ length, $r_0:$ Proximal lumen radius, $E:$ Young's modulus, $R_T:$ Terminal resistance, and $C_T:$ Terminal capacitance.}
\label{tab:vessels}
\begin{tabular}{clccccc} \hline
\# & Artery\tnote{1}& $\ell$  & $r_0$  & $E$  & $R_T$  & $C_T$ \\
  & & (\si{mm})  & (\si{mm})  & (\si{kPa})  & ($10^9$\si{ \pascal\second\per\metre\cubed})  & ($10^{-10}$\si{\metre\cubed\per\pascal}) \\ \hline
01 & Ascending aorta    & \hphantom{0}40 & 12.00            & \hphantom{0}400 &--                & --               \\
02 & Aortic arch I      & \hphantom{0}20 & 11.20            & \hphantom{0}400 &--                & --               \\
03 & Brachiocephalic    & \hphantom{0}34 & \hphantom{0}6.20 & \hphantom{0}400 &--                & --               \\
04 & Aortic arch II     & \hphantom{0}39 & 10.70            & \hphantom{0}400 &--                & --               \\
05 & L CCA              & 208            & \hphantom{0}2.50 & \hphantom{0}400 &--                & --               \\
06 & R CCA              & 177            & \hphantom{0}2.50 & \hphantom{0}400 &--                & --               \\
07 & R Subclavian       & \hphantom{0}34 & \hphantom{0}4.23 & \hphantom{0}400 &--                & --               \\
08 & Thoracic aorta     & 156            & \hphantom{0}9.99 & \hphantom{0}400 & \hphantom{0}0.18 & 38.7\hphantom{0} \\
09 & L Subclavian       & \hphantom{0}34 & \hphantom{0}4.23 & \hphantom{0}400 &--                & --               \\
10/13 & ECA             & 177            & \hphantom{0}1.50 & \hphantom{0}800 & \hphantom{0}5.43 & \hphantom{0}1.27 \\
11/12 & ICA I           & 177            & \hphantom{0}2.00 & \hphantom{0}800 &--                & --               \\
14/17 & Vertebral       & 148            & \hphantom{0}1.36 & \hphantom{0}800 &--                & --               \\
15/16 & Brachial        & 422            & \hphantom{0}4.03 & \hphantom{0}400 & \hphantom{0}2.68 & \hphantom{0}2.58 \\
18/21 & ICA II          & \hphantom{00}5 & \hphantom{0}2.00 & 1600            &--                & --               \\
19/20 & PCoA            & \hphantom{0}15 & \hphantom{0}0.73 & 1600            &--                & --               \\
22 & Basilar I          & \hphantom{0}25 & \hphantom{0}1.62 & 1600            &--                & --               \\
23/24 & MCA             & 119            & \hphantom{0}1.43 & 1600            & \hphantom{0}5.97 & \hphantom{0}1.16 \\
25/26 & ACA I           & \hphantom{0}12 & \hphantom{0}1.17 & 1600            &--                & --               \\
27/28 & PCA I           & \hphantom{00}5 & \hphantom{0}1.07 & 1600            &--                & --               \\
29/30 & ACA II          & 103            & \hphantom{0}1.20 & 1600            & \hphantom{0}8.48 & \hphantom{0}0.82 \\
31 & ACoA               & \hphantom{00}3 & \hphantom{0}0.74 & 1600            &--                & --               \\
32/33 & PCA II          & \hphantom{0}86 & \hphantom{0}1.05 & 1600            & 11.08            & \hphantom{0}0.62 \\
34 & Basilar II         & \hphantom{00}1 & \hphantom{0}1.62 & 1600            &--                & --               \\
35 & Basilar III        & \hphantom{00}3 & \hphantom{0}1.62 & 1600            &--                & --               \\
36/37 & SCA             & \hphantom{0}86 & \hphantom{0}0.65 & 1600            & 25.0\hphantom{0} & \hphantom{0}0.62 \\\hline
\end{tabular}
\begin{tablenotes}
\item[1] Acronyms:Anterior cerebral artery (ACA), Anterior communicating artery (ACoA),
Common carotid artery (CCA), External carotid artery (ECA),
Internal carotid artery (ICA), Middle cerebral artery (MCA),
Posterior cerebral artery (PCA), Posterior communicating artery (PCoA),
Superior cerebellar artery (SCA).
\end{tablenotes}
\end{threeparttable}
}
\end{table}

Lasting one cardiac cycle, the simulated pulsatile blood flow of each vessel must be converted to the electrical property of interest. Visser's model, a nonlinear function that relates blood resistivity changes to the average blood velocity in a cylindrical vessel, can be used for this purpose \citep{Visser1989,Visser1992,Hoetink}
\begin{align}
    \frac{\Delta \rho_{\ell}}{\rho_0}=-0.45  H \left[1- \exp\left(-0.26 \left\vert\frac{\media v}{R}\right\vert^{0.39}\right)\right], \label{eq:visser}
\end{align}
where $\Delta \rho_{\ell}$ is the longitudinal resistivity change to the reference (still blood) resistivity $\rho_0$, $H$ is the hematocrit (volume percentage of red blood cells in the blood), $\media v$ is the average cross-sectional velocity and $R$ is the radius of the vessel. Visser's model presents a similar expression for the conductivity, however no expression was derived for other electrical properties. We will hypothesize the conductivity expression can be applied to the permittivity of blood.

Visser's model applies to blood flowing in a rigid vessel and in a defined orientation. Measurements taken from impedance cardiography studies in humans showed relative variations $\Delta \rho_{\ell}/\rho_0$ smaller (15\% maximum) than predicted from Visser's model in the same conditions (25\% maximum), 60\% reduction \citep{Raaijmakers}. The difference can be explained by the fact that the vessels are not straight and have different orientations.
To accommodate this discrepancy, changes in blood resistivity were scaled to 60\% of Visser's model \eqref{eq:visser}, as reported in the literature.

For each time step, the electrical property of the blood in each main artery is calculated using Visser's model and used to compute the statistics of the atlas at that time instant.

The volume occupied by the main arteries is small compared to the volume of the brain, however its area of influence is considerable. Each artery is responsible for providing blood to specific areas of the brain, known as brain arterial vascular territories \citep{Kim2019}.  Six main supratentorial vascular territories were modelled, (right/left) MCA, ACA, and PCA, also the (right/left) superior cerebellar artery (SCA). The main brain territories can be seen in \figref{fig:cerebralTerritories}. In addition to these, the (right/left) external carotid territories were also included due to the proximity to the electrodes that can impact measurements. The dynamic model does not consider collateral circulation other than the redundancy coming from the circle of Willis.

\image[H]{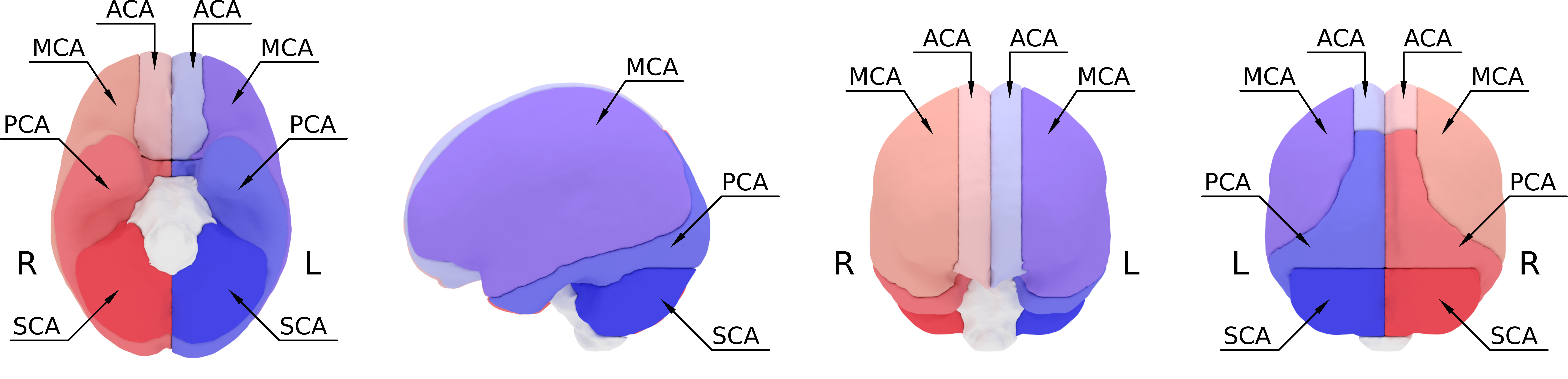}{Main cerebral vascular territories. Acronyms: Anterior cerebral artery (ACA), Middle cerebral artery (MCA), Posterior cerebral artery (PCA), Superior cerebellar artery (SCA).}{fig:cerebralTerritories}{\textwidth}

Blood supply inside each territory is assumed to be proportional to the waveform of the associated main artery. To the best of our knowledge, there are not many studies on electrical property variations of brain tissues along the cardiac cycle. The majority of the studies focus on electrical property changes in response to sensorial or motor activity or epilepsy events \citep{Newell2002,Tidswell2001,Towers2000,Holder1996}. The net electrical property change is caused by a dynamic balance between the amount of blood, extracellular fluids, and cell swelling in a given location and at a given time instant. In this study, the dynamic component of this variation is set to 0.5\% of the main artery of the territory \citep{Tidswell2001}.


\subsubsection{Atlas statistics computation}\label{sec:statisticsB}

Based on the segmentation of the vessels, explained in \secref{sec:segmentationD}, and the location of the vascular territories, presented in \figref{fig:cerebralTerritories}, it is possible to define two characteristic functions per territory, the main vessels in a given territory $\vec{\chi}_m \tamanhoR{N_V}$ and its area of influence $\vec{\chi}^c_m \tamanhoR{N_V}$ (the complement within the vascular territory).

Let $N_M$ be the number of vascular territories. We can write a 3D image of the electrical property of the dynamic component  $\vec{x}(t)\tamanhoC{N_T}$ as
\begin{align}
\vec x(t)  = \sum_{m=1}^{N_M} p_{m,B}(t)\vec{\chi}_m + p_{m,B}^c(t)\vec{\chi}^c_m,
\end{align}
where $p_{m,B}(t)$ and $p_{m,B}^c(t)$ are the electrical properties of blood in the respective segments. The effect of blood in the complements $p_{m,B}^c(t)$ is modelled as a percentage of $p_{m,B}(t)$
\begin{align}
\vec x(t)  = \sum_{m=1}^{N_M} p_{m,B}(t)(\vec{\chi}_m + \alpha_m\vec{\chi}^c_m)=\sum_{m=1}^{N_M} p_{m,B}(t)\vec{\chi}_{\alpha,m},
\end{align}
where $\alpha_m\geq 0$ adjusts the effect. The vector $\vec{\chi}_{\alpha,m}$ represents the influence region of each vascular territory and can be used to compose images as in \eqref{eq:img_rho}. The statistics of the dynamic component is computed following the same procedure presented in \secref{sec:statistics} with the modified characteristic function $\vec{\chi}_{\alpha,m}$.

\section{Results}

\subsection{Static component of the atlas}

\begin{figure}[H]
     \centering
     \begin{subfigure}[b]{\textwidth}
         \centering
         \includegraphics[width=\textwidth]{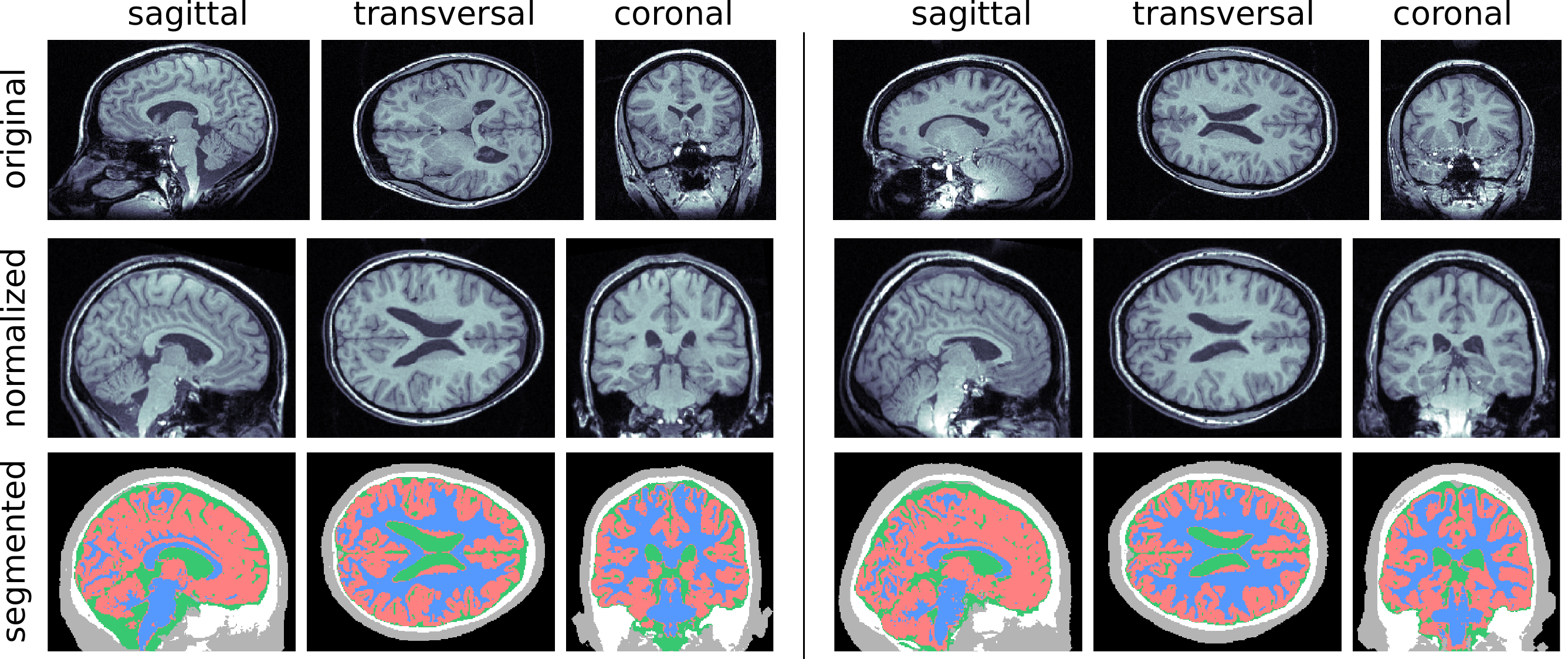}
         \caption{}
         \label{fig:pat03_and_pat04_results}
     \end{subfigure}
     \begin{subfigure}[b]{\textwidth}
         \centering
         \includegraphics[width=0.8\textwidth]{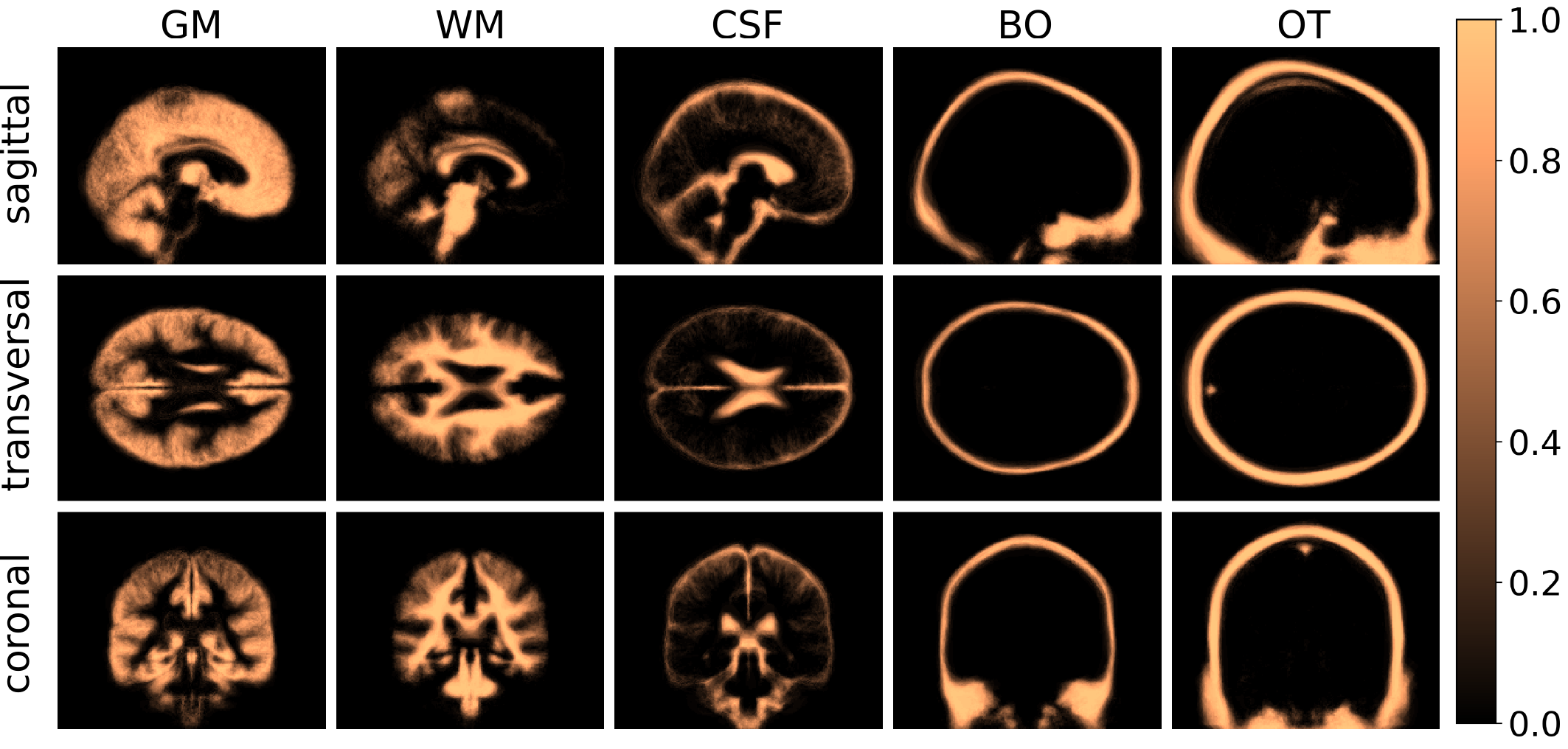}
         \caption{}
         \label{fig:seg_tissues}
     \end{subfigure}
    \caption{Static atlas computations. (a) Processing steps of two representative individuals of the dataset. The first row shows the original images, the second presents the results of the normalization and the third row the result of the segmentation. GM: red, WM: blue, CSF: green, BO: white, and OT: grey. (b) Average of the characteristic functions $\vec{\chi}_t $ for each tissue. }
    \label{fig:three graphs}
\end{figure}

\figref{fig:pat03_and_pat04_results} shows two representative individuals in the segmentation steps to compute the static component of the atlas. The first row contains the original images, the second row contains the normalized images and the third row show the segmented tissues. \figref{fig:seg_tissues} shows the average over the characteristic functions of all individuals and for each segmented tissue. A voxel with a value equal to 1.0 indicates it was classified as the same tissue in all images.

\begin{figure}[H]
     \centering
     \begin{subfigure}[b]{\textwidth}
         \centering
         \includegraphics[width=\textwidth]{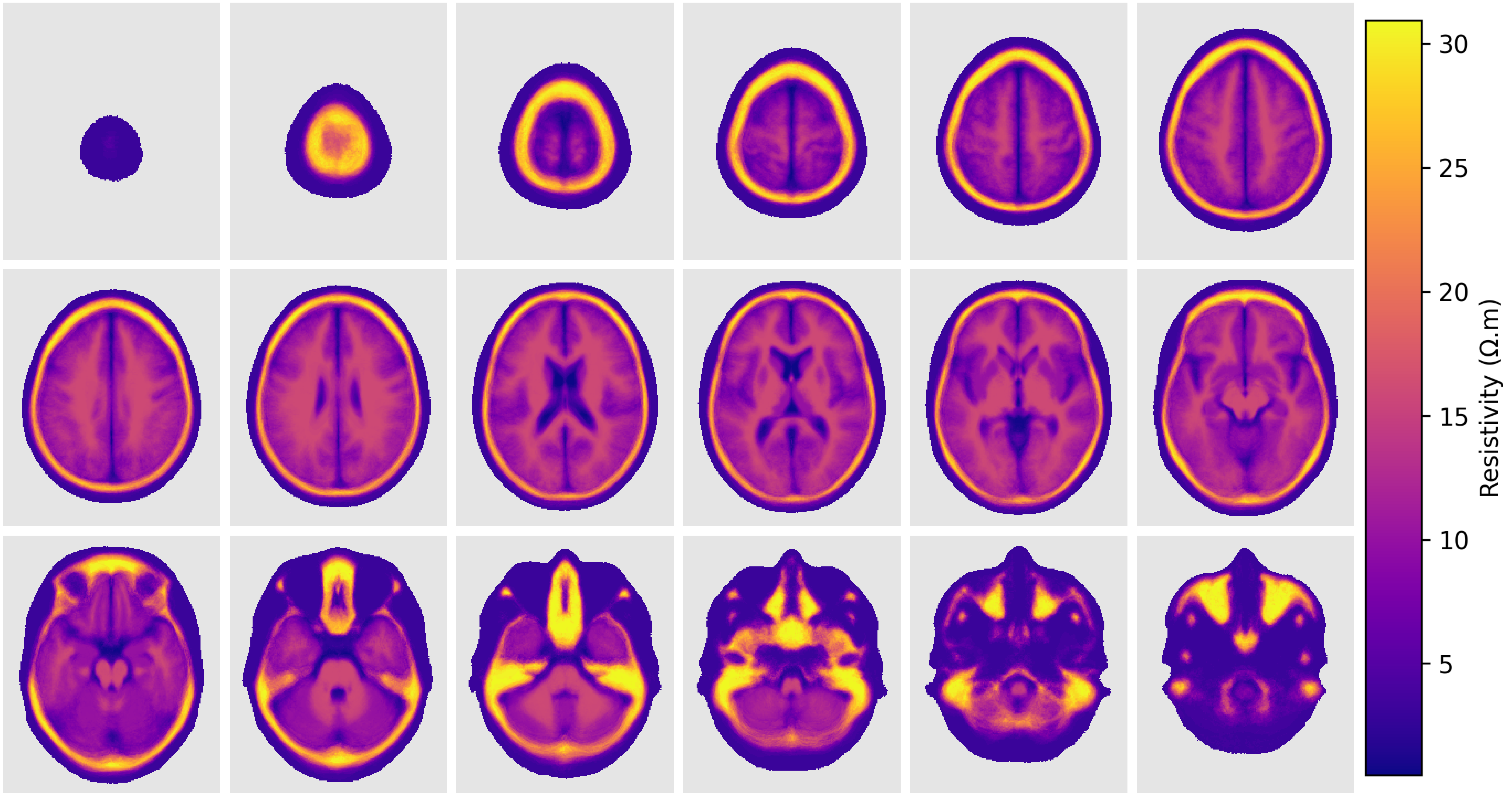}
         \caption{}
         \label{fig:avg_slices}
     \end{subfigure}
     \begin{subfigure}[b]{\textwidth}
         \centering
         \includegraphics[width=\textwidth]{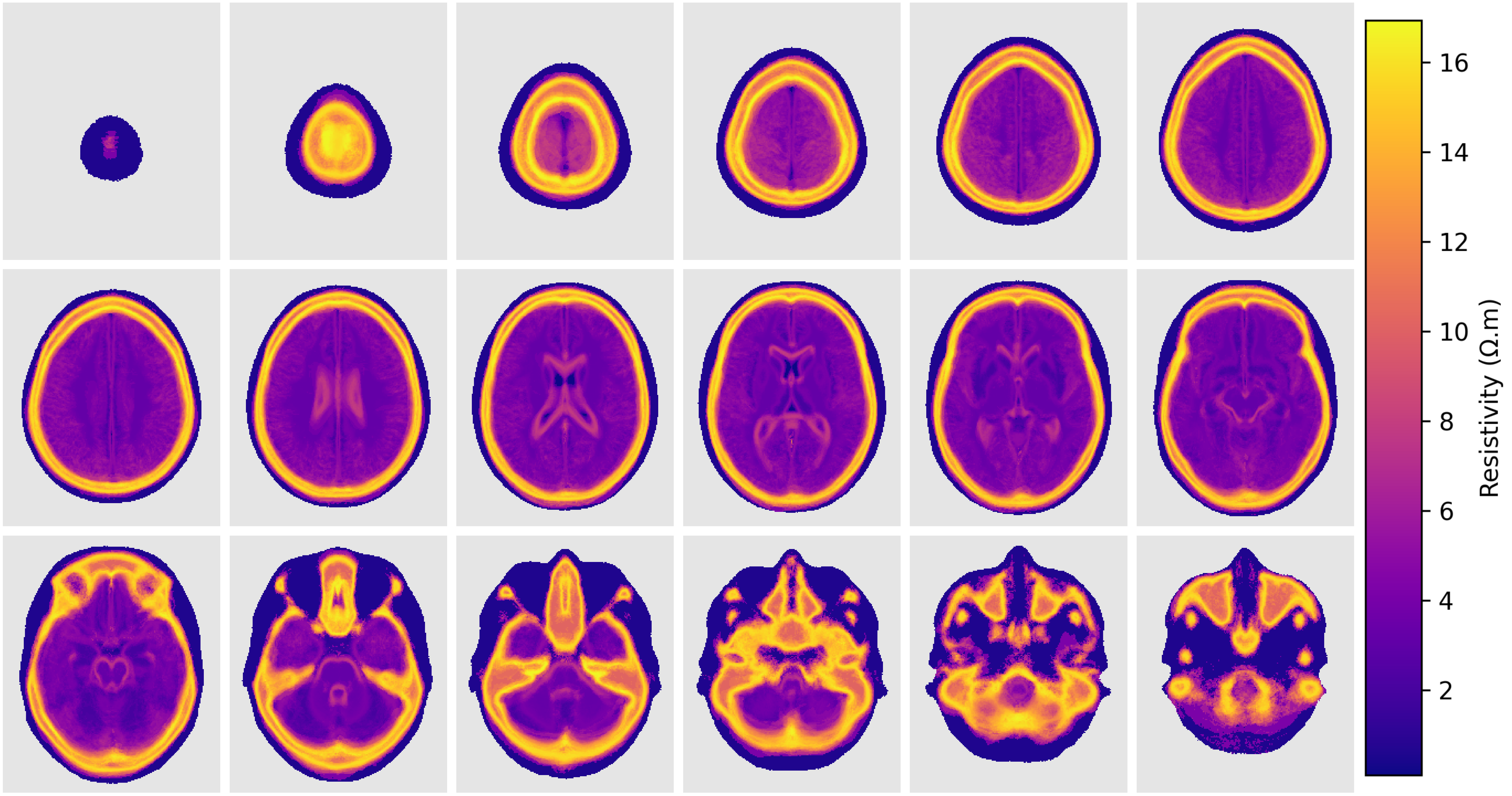}
         \caption{}
         \label{fig:std_slices}
     \end{subfigure}
    \caption{Transversal slices of the static component of the atlas (resistivity) at \SI{1}{\kilo\hertz}. (a) average; (b) standard deviation.}
    \label{fig:atlasResist}
\end{figure}

The static component of the atlas at $\SI{1}{\kilo\hertz}$ is presented in \figref{fig:atlasResist}. The figure shows transversal slices of the average (\figref{fig:avg_slices}) and standard deviation (\figref{fig:std_slices}). It is possible to see high resistivity regions in the forehead, caused by the thick bone and the frontal sinus, in the zygomatic bones and the petrous part of the temporal bone in the base of the skull.

\begin{figure}[H]
     \centering
     \begin{subfigure}[b]{\textwidth}
         \centering
         \includegraphics[width=\textwidth]{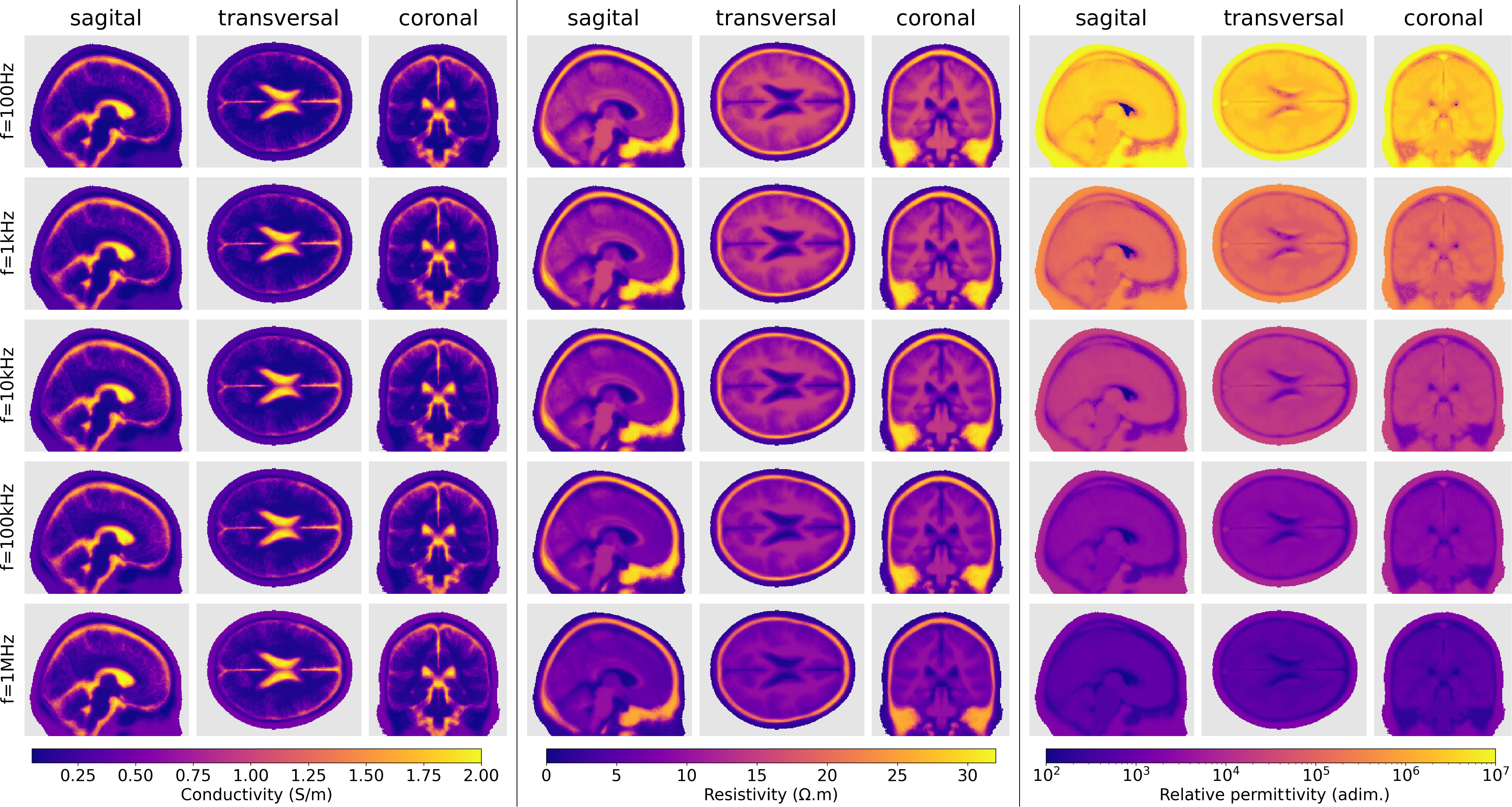}
         \caption{}
         \label{fig:average_sigma_rho_multifreq}
     \end{subfigure}
     \begin{subfigure}[b]{\textwidth}
         \centering
         \includegraphics[width=\textwidth]{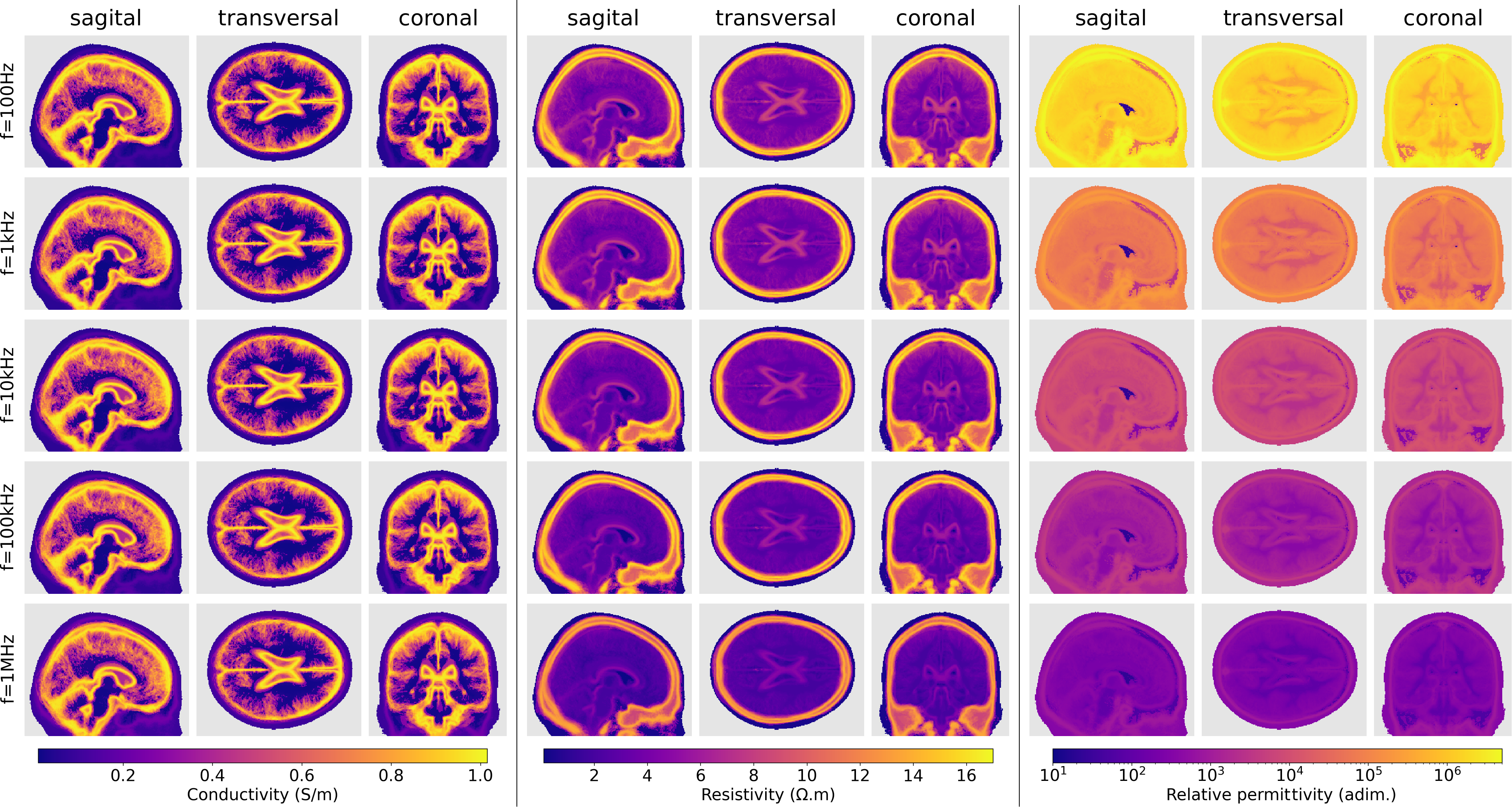}
         \caption{}
         \label{fig:cov_sigma_rho_multifreq}
     \end{subfigure}
    \caption{Statistics of the atlas in different frequencies.  (a) average; (b) standard deviation. }
    \label{fig:atlas_multifreq}
\end{figure}

\figref{fig:atlas_multifreq} shows slices of the atlas built in terms of conductivity, resistivity, and relative permittivity and in different frequencies. The figure shows that the average resistivity and permittivity decrease with increases in frequency while conductivity increases. Although the average process tends to eliminate small features of the images, it is still possible to see small and thin structures inside the brain, like the longitudinal fissure, third and fourth ventricles and central canal.

\subsection{Dynamic component}

The dynamic component of the atlas was computed following the procedure described in \secref{sec:segmentationD}. Transversal slices of the average at $\SI{1}{\kilo\hertz}$ are presented in \figref{fig:avg_slices_vessels}. The main vessels that compose the circle of Willis in the base of the cranial cavity, the dense arterial vascularization in the insular cortex, and the superior sagittal sinus are visible.

\image[H]{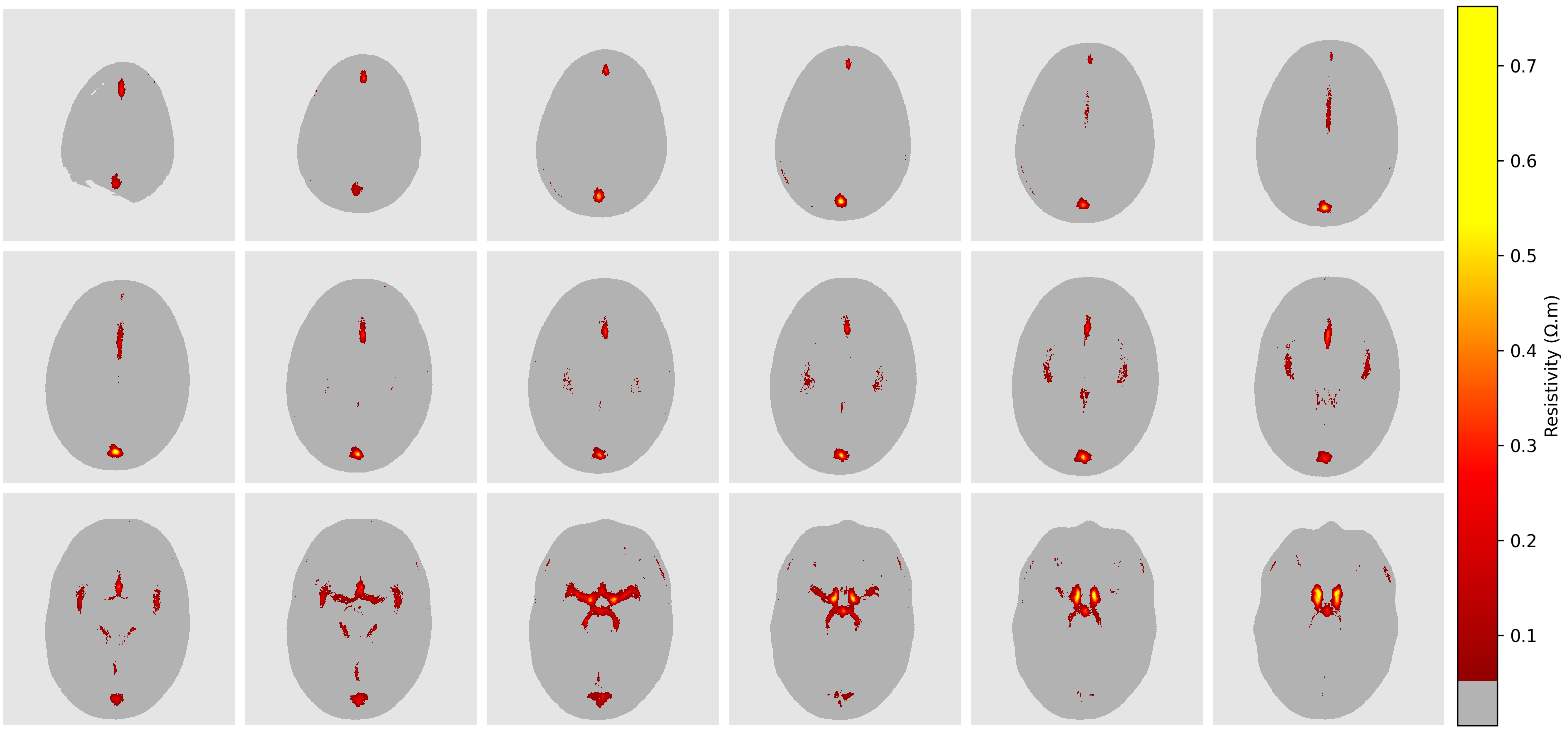}{Transversal slices of the average image of the segmented vessels filled with still blood at \SI{1}{\kilo\hertz}. Small values were masked in grey to emphasize the structure of the main vessels.}{fig:avg_slices_vessels}{\textwidth}

\figref{fig:res_openBF} presents the waveforms obtained from the Navier-Stokes solver simulating one cardiac cycle (60bpm), with hematocrit $H=0.5$. From left to right, the figure presents flow rate, average cross-sectional velocity, static pressure and resistivity changes to still blood following Visser's model \eqref{eq:visser}. Peak velocity occurs approximately \SI{0.25}{\second} after the beginning of the cardiac cycle. Resistivity changes lie within -17\% and -21\% to still blood, indicating the resistivity in these vessels differs considerably from the electrical properties of still blood.

\image[H]{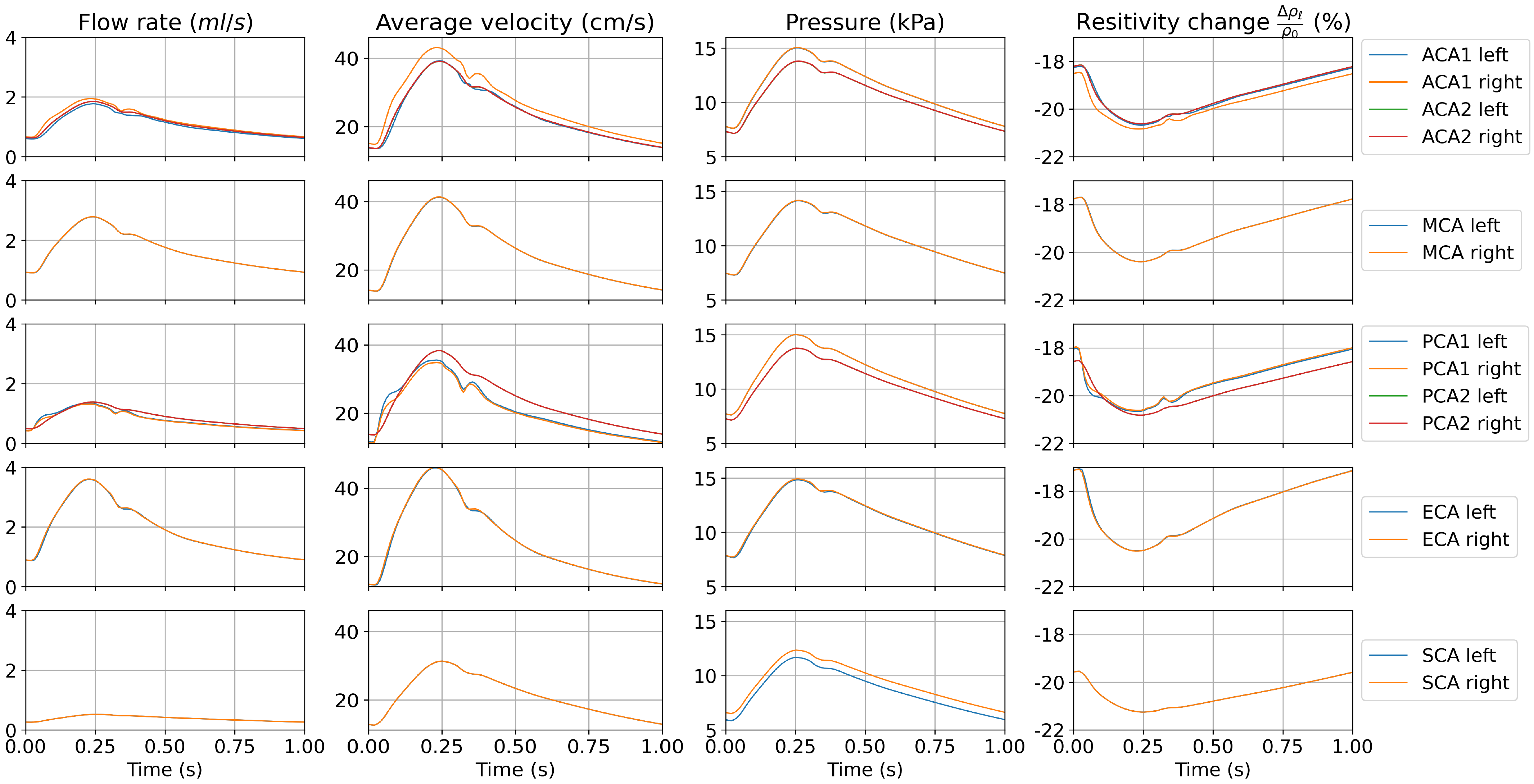}{Waveforms of the main arteries over one cardiac cycle. From left to right: average cross-sectional velocity, static pressure and resistivity changes.}{fig:res_openBF}{\textwidth}

\subsection{Effects of the cardiac cycle on surface measurements for electrical impedance tomography}\label{sec:exampleEIT}

As one example of application, the atlas was employed to simulate EIT surface electrode measurements at $\SI{1}{\kilo\hertz}$ during one cardiac cycle (60bpm) using the finite element method to solve the complete electrode model for  EIT \citep{CHENEY-1999,2005holder}. This type of \textit{in silico} study is important to investigate the possibility of monitoring blood perfusion anomalies in patients.

A segmented head image of an average young adult\footnote{https://www.pedeheadmod.net/pediatric-head-atlases/} was selected to create the geometry \citep{turovets,turovets2,turovets3}. This geometry is not among those used to create the atlas, avoiding statistical biases. The boundary surfaces of the segments were extracted and cleaned to remove artefacts.

\begin{figure}[ht]
     \centering
     \begin{subfigure}[b]{\textwidth}
         \centering
         \includegraphics[width=0.7\textwidth]{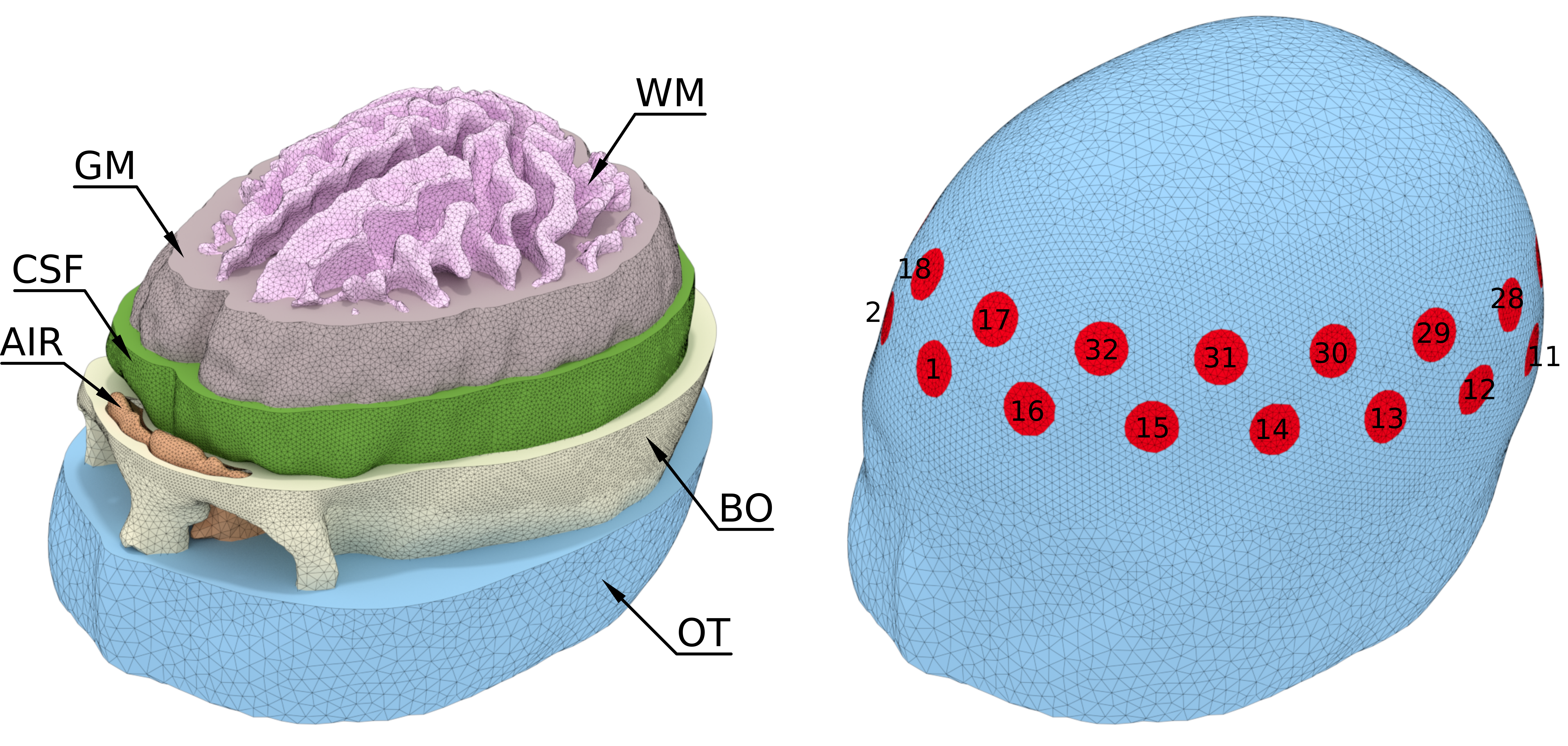}
         \caption{}
         \label{fig:mesh_FEM}
     \end{subfigure}
     \begin{subfigure}[b]{\textwidth}
         \centering
         \includegraphics[width=\textwidth]{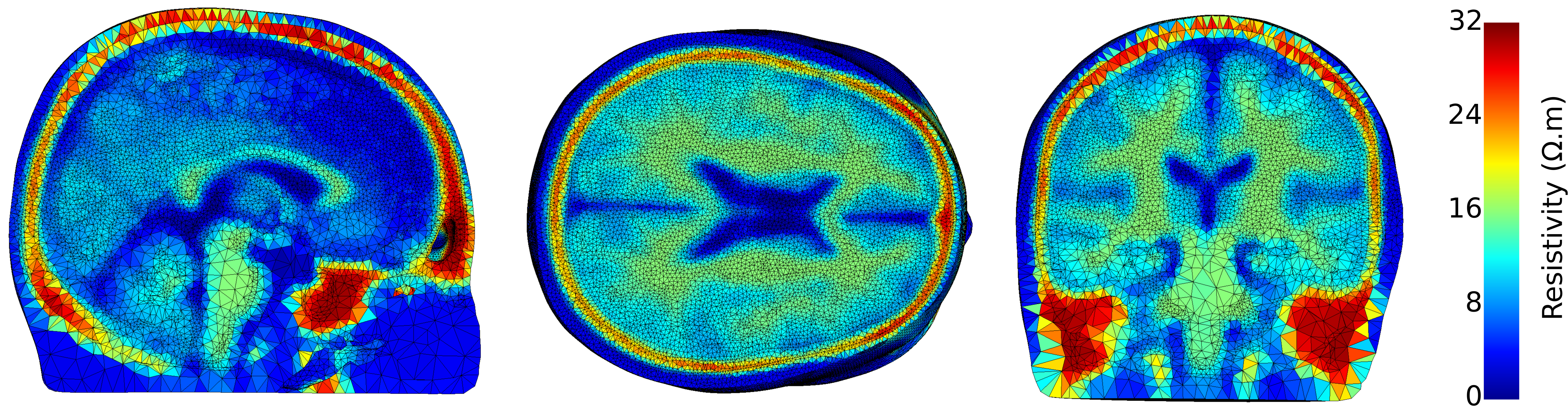}
         \caption{}
         \label{fig:atlas_projection}
     \end{subfigure}
    \caption{Finite element mesh used to simulate EIT measurements. (a) Internal structure and electrode locations; (b) Slices of the projected atlas into the FEM mesh. }
    \label{fig:mesh_Rob}
\end{figure}

A FEM mesh was created using Gmsh software \citep{GMSH2009}. The mesh, presented in \figref{fig:mesh_FEM}, is composed of 2.06 million linear tetrahedral elements, split into six segments and 32 electrodes with a diameter of $\SI{15}{\milli\metre}$ were placed in two parallel planes with \SI{20}{\milli\metre } of separation. Electrode numbers can be seen in the figure. 

The atlas was projected into the FEM mesh using the symmetric image normalization (SyN) \citep{AVANTS2008}. The method requires two 3D binary images with the characteristic functions of the same volume, one in the atlas reference system and one in the FEM mesh reference system. The volumes used for the normalization comprise the soft tissues inside the cranial cavity (GM+WM+CSF).

For the atlas, the characteristic function of the cranial cavity $\vec \chi_C$ is found by taking the average of the sum of the characteristic functions of these tissues over all individuals that compose the atlas, followed by thresholding at 75\%
\begin{align}
\media \chi &= E\{\vec{\chi}_{WM} + \vec{\chi}_{GM} + \vec{\chi}_{CSF}\}\\
    \vec \chi_C(i,j,k) &= 
    \begin{cases}
      1 & \text{if } \media \chi(i,j,k)\geq0.75\\
      0 & \text{otherwise}
    \end{cases},
\end{align}
where $(i,j,k)$ is the coordinates of each voxel. For the FEM mesh, a 3D image can be created from the segmented internal structures of the mesh, seen in \figref{fig:mesh_FEM}, by defining a 3D grid of points that encloses the head and checking if each voxel belongs to the segment GM+WM+CSF.

The affine transformation resulting from the normalization was applied to all voxels of the atlas, projecting them into the FEM mesh reference system. Finally, the values of the atlas were interpolated into the centroids of the tetrahedron. The projection can be seen in \figref{fig:atlas_projection}.

EIT measurement simulation was performed by imposing sinusoidal bipolar current injection of $\SI{1}{\milli\ampere}$ at $\SI{1}{\kilo\hertz}$ and computing the electrode voltage measurements. Current pattern follows a skip-8 scheme to allow diametral current injection (two planes with 16 electrodes each) \citep{LuppiSilva2017}. This choice mitigates the electrical shunting effect of the skull that causes the majority of the current to flow along the scalp only if the pair of injecting electrodes are too close. The simulated measurements are presented in \figref{fig:measurement_07} in four time instants along the cardiac cycle.

\image[H]{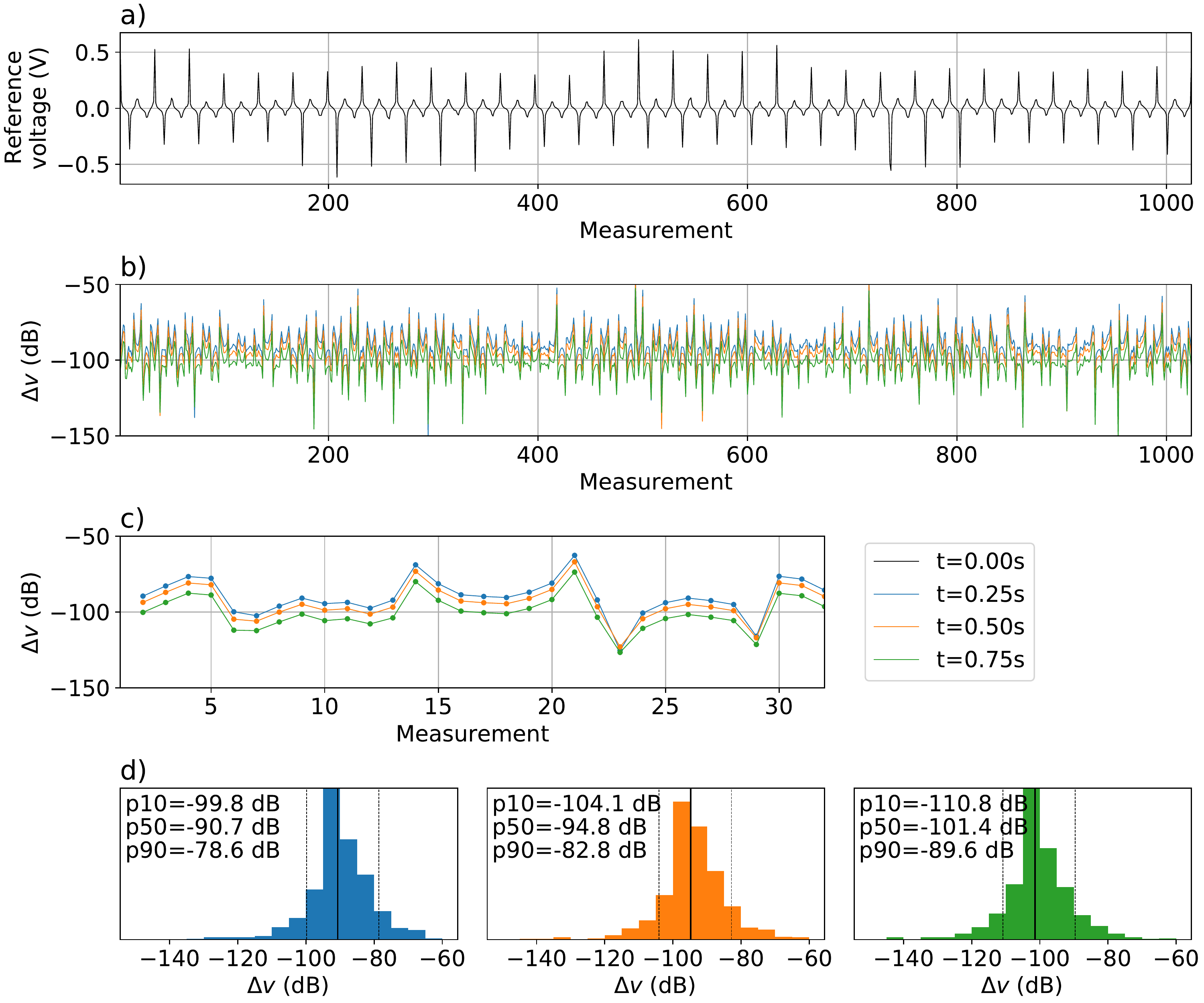}{Simulated electrode measurements. (a) reference measurements $\vec v_0$; (b) relative differences $\Delta \vec v(t)$; (c) first 32 relative differences $\Delta \vec v(t)$; (d) histograms of the differences $\Delta \vec v(t)$.}{fig:measurement_07}{0.9\textwidth}

\subfigref{fig:measurement_07}{a} presents the measurements at $t=\SI{0}{\second}$, used as reference measurement $\vec v_0$. \subfigref{fig:measurement_07}{b} shows relative differences $\Delta \vec v(t)$, in dB, between measurements in three other time instants $\vec v(t)$ and the reference $\vec v_0$, defined as
\begin{align}
    \Delta \vec v(t)=20\log_{10}\left(\left \vert\frac{\vec v(t)-\vec v_0}{\vec v_0}\right\vert\right),
\end{align}
where the division is computed element-wise. \subfigref{fig:measurement_07}{c} is a plot of the same differences for the first 32 measurements and show that the largest differences are measured at $t=\SI{0.25}{\second}$, time instant when we have peak velocities, as presented before in \figref{fig:res_openBF}. \subfigref{fig:measurement_07}{d} presents histograms of the differences $\Delta \vec v(t)$. In the same histograms, the vertical lines represent the percentiles 10\%, 50\% (median), and 90\% together with their numeric values (p10, p50, and p90). 

\section{Discussion}

Conductor volume problems in electroencephalography and electrical bioimpedance cerebral monitoring require a 3D model of the head and its electrical properties for solving the associated PDEs numerically. In many situations, a 3D model of the head of the patient is not available or an average head model is preferred. Even in cases when the model is available via MRI or CT images, the electrical properties of the tissues might not be completely known due to natural variability. This work presents a novel 4D (3D+T) statistical anatomical atlas of the electrical properties of the human head for electrophysiology applications.

The atlas was built for an average head shape and can be normalized to specific geometries. This process was exemplified with one EIT application. The normalization step optimized the alignment of the cranial cavity, volume comprising GM WM and CSF segments. In our experiments, this choice produces a better match between the atlas and the FEM mesh. However, this choice can cause small artefacts in the external surface of the mesh due to the small thickness of the scalp and its proximity to the skull. These artefacts can be seen in \figref{fig:atlas_projection} where the resistivity near the top of the cranial vault and the anterior part of the frontal bone affect the resistivity near the external surface.


The electrical properties of biological tissues are frequency dependent. The atlas can be built for different frequencies as exemplified in \figref{fig:atlas_multifreq}. This flexibility expands the applicability of the atlas, such as in multifrequency EIT \citep{Horesh,Malone2014}.

Cerebral circulation was also modelled and added to the atlas. The atlas is capable of simulating the pulsatile blood flow in the main cerebral arteries and vascular territories and their effects on electrical measurements. As one example of application, the atlas was employed in an \textit{in silico} study to investigate the possibility of monitoring blood perfusion using EIT. Among other objectives, this type of study is important to provide information on measurement sensitivity necessary to detect perfusion anomalies and serve as a guide for future EIT equipment developments for these applications.  \figref{fig:measurement_07}{d} predicts that EIT equipments need a signal-to-noise ratio between 100 and \SI{125}{\decibel} to identify changes due to the cardiac cycle. This is in good agreement with a previous study \citep{Towers2000} on clamped carotid arteries.

The statistical nature of the atlas also allows quantifying its uncertainty. \figref{fig:atlasResist} shows that the regions with the largest standard variations are located along the boundary of the bones. This can be explained by the fact that small anatomical differences between the skulls of the individuals cause large resistivity variations due to the difference between the resistivity of bones and other tissues around them.

The proposed atlas has some limitations. It is known that ageing increases the stiffness of the vessels, however the atlas does not include ageing effects on the stiffness of the walls of the arteries. Nevertheless, the atlas can be adjusted by setting Young's modulus of the vessels accordingly with the age of the population. Except for the redundancy caused by the circle of Willis, collateral brain circulation was not modelled either caused by preexisting vascular redundancy or neovascularization. The venous side of the circulation was not modelled. Although only five main tissues were segmented, the method can be readily extended to accommodate more tissues. All tissues were modelled as isotropic, even though it is known that some tissues are anisotropic. Extending the atlas to anisotropic tissue is possible but increases the complexity substantially. Also, the scalp was modelled as a uniform tissue, however it is a multi-layer tissue, composed of skin, connective tissue, epicranial aponeurosis, and muscles that have different electrical properties. Due to its proximity to the electrodes, scalp mismodellings can impact EIT recovered images. Finally, the dynamic effect of blood circulation in each territory was modelled as proportional to the velocity of the blood in the main vessel. This approximation does not take into consideration variations in the volume of blood in a given region, for example when the brain responds to external stimuli. The current limitations of the atlas act as motivation for future research topics. These challenging limitation will be the focus of future works to further improve the anatomical atlas.

\section{Conclusion}

We presented a novel anatomical atlas of the electrical properties of the human head. To the best of our knowledge, the present model is the first model capable of simulating cerebral circulation and its effects on electrical measurements.
Despite the limitations, the atlas brings important implications to cerebral electrophysiology studies.  This novelty has the potential to become an important tool for \textit{in silico} studies on cerebral circulation and electrophysiology, such as electrical measurement sensitivity to vascular pathologic conditions like stroke classification and monitoring, arterial vasospasms, and arteriovenous malformation. The atlas can also be used as statistical prior information for inverse problems in EEG and EIT and to create training sets for machine learning algorithms.

\section*{Acknowledgements}

The work was funded in part by the Jane and Aatos Erkko Foundation, project “Electrical Impedance Tomography---a novel method for improved diagnosis of stroke”, the Academy of Finland (Centre of Excellence in Inverse Modelling and Imaging, decision number  312339), the São Paulo Research Foundation--FAPESP (Process numbers: 2019/09154-7 and 2017/18378-0) and Coordenação de Aperfeiçoamento de Pessoal de Nível Superior (CAPES) -- Finance Code 001.

The MR brain images from healthy volunteers used in this paper were collected and made available by the CASILab at The University of North Carolina at Chapel Hill and were distributed by the MIDAS Data Server at Kitware, Inc. (\url{https://www.insight-journal.org/midas/community/view/21})

\bibliographystyle{./dcu.bst}
\bibliography{main.bib}

@article{agnelli2020classification,
  title={Classification of stroke using neural networks in electrical impedance tomography},
  author={Agnelli, Juan Pablo and {\c{C}}{\"o}l, Aynur and Lassas, Matti and Murthy, Rashmi and Santacesaria, Matteo and Siltanen, Samuli},
  journal={Inverse Problems},
  volume={36},
  number={11},
  pages={115008},
  year={2020},
  publisher={IOP Publishing}
}

@article{candiani2019computational,
  title={Computational framework for applying electrical impedance tomography to head imaging},
  author={Candiani, Valentina and Hannukainen, Antti and Hyvonen, Nuutti},
  journal={SIAM Journal on Scientific Computing},
  volume={41},
  number={5},
  pages={B1034--60},
  year={2019},
  publisher={SIAM}
}

@article{candiani2020neural,
  title={Neural networks for classification of strokes in electrical impedance tomography on a 3D head model},
  author={Candiani, Valentina and Santacesaria, Matteo},
  journal={arXiv preprint arXiv:2011.02852},
  year={2020}, 
  pages = {1--20}
}

@Article{Dowrick2016,
  author    = {T Dowrick and C Blochet and D Holder},
  title     = {In vivo bioimpedance changes during haemorrhagic and ischaemic stroke in rats: towards 3D stroke imaging using electrical impedance tomography},
  journal   = {Physiological Measurement},
  year      = {2016},
  volume    = {37},
  number    = {6},
  pages     = {765--84},
  month     = {may},
  doi       = {10.1088/0967-3334/37/6/765},
  publisher = {{IOP} Publishing},
}

@Article{Song2018,
  author    = {Jiali Song and Rongqing Chen and Lin Yang and Ge Zhang and Weichen Li and Zhanqi Zhao and Canhua Xu and Xiuzhen Dong and Feng Fu},
  title     = {Electrical Impedance Changes at Different Phases of Cerebral Edema in Rats with Ischemic Brain Injury},
  journal   = {{BioMed} Research International},
  year      = {2018},
  volume    = {2018},
  pages     = {1--10},
  month     = {jun},
  doi       = {10.1155/2018/9765174},
  publisher = {Hindawi Limited},
}

@Article{Proena2020,
  author    = {Martin Proen{\c{c}}a and Fabian Braun and Mathieu Lemay and Josep Sol{\`{a}} and Andy Adler and Thomas Riedel and Franz H. Messerli and Jean-Philippe Thiran and Stefano F. Rimoldi and Emrush Rexhaj},
  title     = {Non-invasive pulmonary artery pressure estimation by electrical impedance tomography in a controlled hypoxemia study in healthy subjects},
  journal   = {Scientific Reports},
  year      = {2020},
  volume    = {10},
  number    = {1},
  month     = {dec},
  doi       = {10.1038/s41598-020-78535-4},
  publisher = {Springer Science and Business Media {LLC}},
}

@InCollection{SPM,
  title     = {Chapter 31 - Experimental Design and Statistical Parametric Mapping},
  booktitle = {Human Brain Function (Second Edition)},
  publisher = {Academic Press},
  year      = {2004},
  author    = {Richard S.J. Frackowiak and Karl J. Friston and Christopher D. Frith and Raymond J. Dolan and Cathy J. Price and Semir Zeki and John T. Ashburner and William D. Penny},
  pages     = {599-632},
  address   = {Burlington},
  edition   = {Second Edition},
  isbn      = {978-0-12-264841-0},
  doi       = {https://doi.org/10.1016/B978-012264841-0/50033-0},
}

@InCollection{Min2019,
  author    = {Mart Min and Hip K{\~{o}}iv and Eiko Priidel and Ksenija Pesti and Paul Annus},
  title     = {Noninvasive Acquisition of the Aortic Blood Pressure Waveform},
  booktitle = {Wearable Devices - the Big Wave of Innovation},
  publisher = {{IntechOpen}},
  year      = {2019},
  month     = {dec},
  doi       = {10.5772/intechopen.86065},
}

@article{Newell2002,
  doi = {10.1088/0967-3334/23/1/321},
  year = {2002},
  month = jan,
  publisher = {{IOP} Publishing},
  volume = {23},
  number = {1},
  pages = {203--9},
  author = {J C Newell and R S Blue and D Isaacson and G J Saulnier and A S Ross},
  title = {Phasic three-dimensional impedance imaging of cardiac activity},
  journal = {Physiological Measurement}
}

@article{Malone2014,
  doi = {10.1088/0967-3334/35/6/1051},
  year = {2014},
  month = may,
  publisher = {{IOP} Publishing},
  volume = {35},
  number = {6},
  pages = {1051--66},
  author = {Emma Malone and Markus Jehl and Simon Arridge and Timo Betcke and David Holder},
  title = {Stroke type differentiation using spectrally constrained multifrequency {EIT}: evaluation of feasibility in a realistic head model},
  journal = {Physiological Measurement}
}

@article{Tidswell2001,
  doi = {10.1006/nimg.2000.0698},
  year = {2001},
  month = feb,
  publisher = {Elsevier {BV}},
  volume = {13},
  number = {2},
  pages = {283--94},
  author = {Tom Tidswell and Adam Gibson and Richard H. Bayford and David S. Holder},
  title = {Three-Dimensional Electrical Impedance Tomography of Human Brain Activity},
  journal = {{NeuroImage}}
}

@article{Towers2000,
  doi = {10.1088/0967-3334/21/1/315},
  year = {2000},
  month = feb,
  publisher = {{IOP} Publishing},
  volume = {21},
  number = {1},
  pages = {119--24},
  author = {C M Towers and H McCann and M Wang and P C Beatty and C J D Pomfrett and M S Beck},
  title = {3D simulation of {EIT} for monitoring impedance variations within the human head},
  journal = {Physiological Measurement}
}

@article{Holder1996,
  doi = {10.1088/0967-3334/17/4a/022},
  year = {1996},
  month = nov,
  publisher = {{IOP} Publishing},
  volume = {17},
  number = {4A},
  pages = {A179--86},
  author = {D S Holder and A Rao and Y Hanquan},
  title = {Imaging of physiologically evoked responses by electrical impedance tomography with cortical electrodes in the anaesthetized rabbit},
  journal = {Physiological Measurement}
}

@article{LuppiSilva2017,
  doi = {10.1016/j.conengprac.2016.03.003},
  year = {2017},
  month = jan,
  publisher = {Elsevier {BV}},
  volume = {58},
  pages = {276--86},
  author = {Olavo Luppi Silva and Raul Gonzalez Lima and Thiago Castro Martins and Fernando Silva Moura and Renato Seiji Tavares and Marcos Sales Guerra Tsuzuki},
  title = {Influence of current injection pattern and electric potential measurement strategies in electrical impedance tomography},
  journal = {Control Engineering Practice}
}

@Book{2005holder,
  title     = {Electrical impedance tomography: Methods, History and Applications},
  publisher = {IOP Publishing Ltd},
  year      = {2005},
  author    = {David S. Holder},
  address   = {Cornwall, UK},
  edition   = {1},
}

@ARTICLE{CHENEY-1999,
  author = {M. Cheney and D. Isaacson and J. C. Newell},
  title = {Electrical Impedance Tomography},
  journal = {SIAM Review},
  year = {1999},
  volume = {41},
  pages = {85--101},
  number = {1},
}

@ARTICLE{Frerichs2016,
  author    = {In{\'{e}}z Frerichs and Marcelo B P Amato and Anton H van Kaam and David G Tingay and Zhanqi Zhao and Bart{\l}omiej Grychtol and Marc Bodenstein and Herv{\'{e}} Gagnon and Stephan H B\"{o}hm and Eckhard Teschner and Ola Stenqvist and Tommaso Mauri and Vinicius Torsani and Luigi Camporota and Andreas Schibler and Gerhard K Wolf and Diederik Gommers and Steffen Leonhardt and Andy Adler},
  title     = {Chest electrical impedance tomography examination, data analysis, terminology, clinical use and recommendations: consensus statement of the {TRanslational} {EIT} {developmeNt} {stuDy} group},
  journal   = {Thorax},
  year      = {2016},
  volume    = {72},
  number    = {1},
  pages     = {83--93},
  month     = {sep},
  doi       = {10.1136/thoraxjnl-2016-208357},
  publisher = {{BMJ}},
}

@InCollection{Adler2019,
  author    = {Andy Adler and Alistair Boyle},
  title     = {Electrical Impedance Tomography},
  booktitle = {Wiley Encyclopedia of Electrical and Electronics Engineering},
  publisher = {John Wiley $\&$ Sons, Inc},
  year      = {2019},
  editor    = {John G. Webster},
  pages     = {1--16},
  address   = {New Jersey, United States},
  edition   = {2},
  month     = {5},
  doi       = {10.1002/047134608x.w1431.pub2},
}

@Article{MARTINS2019442,
  author    = {T. C. Martins and André Kubagawa Sato and Fernando Silva Moura and Erick Dario León Bueno de Camargo and Olavo Luppi Silva and Talles Batista Rattis Santos and Zhanqi Zhao and Knut M\"oeller and Marcelo Brito Passos Amato and Jennifer L. Mueller and Raul Gonzalez Lima and Marcos S. G. Tsuzuki},
  title     = {A review of electrical impedance tomography in lung applications: Theory and algorithms for absolute images},
  journal   = {Annual Reviews in Control},
  year      = {2019},
  volume    = {48},
  pages     = {442 - 71},
  issn      = {1367-5788},
  doi       = {https://doi.org/10.1016/j.arcontrol.2019.05.002},
}

@Article{nipype,
  author    = {Gorgolewski, Krzysztof and Burns, Christopher and Madison, Cindee and Clark, Dav and Halchenko, Yaroslav and Waskom, Michael and Ghosh, Satrajit},
  title     = {Nipype: A Flexible, Lightweight and Extensible Neuroimaging Data Processing Framework in Python},
  journal   = {Frontiers in Neuroinformatics},
  year      = {2011},
  volume    = {5},
  pages     = {13},
  issn      = {1662-5196},
  doi       = {10.3389/fninf.2011.00013},
}

@Article{Michel2019,
  author    = {Christoph M. Michel and Denis Brunet},
  title     = {{EEG} Source Imaging: A Practical Review of the Analysis Steps},
  journal   = {Frontiers in Neurology},
  year      = {2019},
  PAGES={325},  
  volume    = {10},
  month     = {apr},
  doi       = {10.3389/fneur.2019.00325},
  publisher = {Frontiers Media {SA}},
}

@Article{Cho2015,
  author    = {Jae-Hyun Cho and Johannes Vorwerk and Carsten H. Wolters and Thomas R. Kn\"{o}sche},
  title     = {Influence of the head model on {EEG} and {MEG} source connectivity analyses},
  journal   = {{NeuroImage}},
  year      = {2015},
  volume    = {110},
  pages     = {60--77},
  month     = {apr},
  doi       = {10.1016/j.neuroimage.2015.01.043},
  publisher = {Elsevier {BV}},
}

@Article{Vorwerk2014,
  author    = {Johannes Vorwerk and Jae-Hyun Cho and Stefan Rampp and Hajo Hamer and Thomas R. Kn\"{o}sche and Carsten H. Wolters},
  title     = {A guideline for head volume conductor modeling in {EEG} and {MEG}},
  journal   = {{NeuroImage}},
  year      = {2014},
  volume    = {100},
  pages     = {590--607},
  month     = {oct},
  doi       = {10.1016/j.neuroimage.2014.06.040},
  publisher = {Elsevier {BV}},
}

@Article{AkalinAcar2013,
  author    = {Zeynep Akalin Acar and Scott Makeig},
  title     = {Effects of Forward Model Errors on {EEG} Source Localization},
  journal   = {Brain Topography},
  year      = {2013},
  volume    = {26},
  number    = {3},
  pages     = {378--96},
  month     = {jan},
  doi       = {10.1007/s10548-012-0274-6},
  publisher = {Springer Science and Business Media {LLC}},
}

@Article{Hallez2007,
  author    = {Hallez, Hans and Vanrumste, Bart and Grech, Roberta and Muscat, Joseph and De Clercq, Wim and Vergult, Anneleen and D'Asseler, Yves and Camilleri, Kenneth P. and Fabri, Simon G. and Van Huffel, Sabine and Lemahieu, Ignace},
  title     = {Review on solving the forward problem in EEG source analysis},
  journal   = {Journal of NeuroEngineering and Rehabilitation},
  year      = {2007},
  volume    = {4},
  number    = {1},
  pages     = {46},
  month     = {Nov},
  issn      = {1743-0003},
  day       = {30},
  doi       = {10.1186/1743-0003-4-46},
}

@Article{Grech2008,
  author    = {Grech, Roberta and Cassar, Tracey and Muscat, Joseph and Camilleri, Kenneth P. and Fabri, Simon G. and Zervakis, Michalis and Xanthopoulos, Petros and Sakkalis, Vangelis and Vanrumste, Bart},
  title     = {Review on solving the inverse problem in EEG source analysis},
  journal   = {Journal of NeuroEngineering and Rehabilitation},
  year      = {2008},
  volume    = {5},
  number    = {1},
  pages     = {25},
  month     = {Nov},
  issn      = {1743-0003},
  day       = {07},
  doi       = {10.1186/1743-0003-5-25},
}

@Book{jatoi2017brain,
  title     = {Brain source localization using EEG signal analysis},
  publisher = {CRC Press},
  year      = {2017},
  author    = {Munsif Ali Jatoi and Nidal Kamel},
  address   = {Boca Raton, FL},
  isbn      = {978-1-4987-9934-8},
}

@Book{kaipio2005statistical,
  title     = {Statistical and computational inverse problems},
  publisher = {Springer},
  year      = {2005},
  author    = {Jari Kaipio and Erkki Somersalo},
  address   = {New York},
  isbn      = {978-0387220734},
}

@InBook{Bradac2017,
  pages     = {105--7},
  title     = {Vascular Territories},
  publisher = {Springer International Publishing},
  year      = {2017},
  author    = {Bradac, Gianni Boris},
  address   = {Cham},
  isbn      = {978-3-319-57228-4},
  booktitle = {Applied Cerebral Angiography: Normal Anatomy and Vascular Pathology},
  doi       = {10.1007/978-3-319-57228-4_8},
}

@Article{circulation1,
  author    = {Chandra, Ankush. and Li, William. and Stone, Christopher. and Geng, Xiaokun. and Ding, Yuchuan.},
  title     = {{The cerebral circulation and cerebrovascular disease I: Anatomy}},
  journal   = {Brain Circulation},
  year      = {2017},
  volume    = {3},
  number    = {2},
  pages     = {45--56},
  doi       = {10.4103/bc.bc_10_17},
}

@Article{Kim2019,
  author    = {Dong-Eog Kim and Jinseong Jang and Dawid Schellingerhout and Wi-Sun Ryu and Jong-Ho Park and Su-Kyoung Lee and Dongmin Kim and Hee-Joon Bae},
  title     = {Supratentorial Cerebral Arterial Territories for Computed Tomograms: A Mapping Study in 1160 Large Artery Infarcts},
  journal   = {Scientific Reports},
  year      = {2019},
  volume    = {9},
  number    = {1},
  month     = {aug},
  doi       = {10.1038/s41598-019-48266-2},
  publisher = {Springer Science and Business Media {LLC}},
}

@Article{Sakka2011,
  author    = {L. Sakka and G. Coll and J. Chazal},
  title     = {Anatomy and physiology of cerebrospinal fluid},
  journal   = {European Annals of Otorhinolaryngology, Head and Neck Diseases},
  year      = {2011},
  volume    = {128},
  number    = {6},
  pages     = {309--16},
  month     = {dec},
  doi       = {10.1016/j.anorl.2011.03.002},
  publisher = {Elsevier {BV}},
}

@Article{DRIVER_2020,
  author    = {Driver, Ian D. and Traat, Maarika and Fasano, Fabrizio and Wise, Richard G.},
  title     = {Most Small Cerebral Cortical Veins Demonstrate Significant Flow Pulsatility: A Human Phase Contrast MRI Study at 7T},
  journal   = {Frontiers in Neuroscience},
  year      = {2020},
  volume    = {14},
  pages     = {415},
  issn      = {1662-453X},
  doi       = {10.3389/fnins.2020.00415},
}

@Article{gabriel1,
  author        = {C Gabriel and S Gabriel and E Corthout},
  title         = {The dielectric properties of biological tissues: I. Literature survey},
  journal       = {Physics in Medicine $\&$ Biology},
  year          = {1996},
  volume        = {41},
  number        = {11},
  pages         = {2231--49},
  urlaccessdate = {01.02.2019},
}

@Article{gabriel2,
  author        = {S Gabriel and R W Lau and C Gabriel},
  title         = {The dielectric properties of biological tissues: II. Measurements in the frequency range 10 Hz to 20 GHz},
  journal       = {Physics in Medicine $\&$ Biology},
  year          = {1996},
  volume        = {41},
  number        = {11},
  pages         = {2251--69},
  urlaccessdate = {01.02.2019},
}

@Article{gabriel3,
  author        = {S Gabriel and R W Lau and C Gabriel},
  title         = {The dielectric properties of biological tissues: III. Parametric models for the dielectric spectrum of tissues},
  journal       = {Physics in Medicine $\&$ Biology},
  year          = {1996},
  volume        = {41},
  number        = {11},
  pages         = {2271--93},
  urlaccessdate = {01.02.2019},
}

@Misc{Andreuccetti,
  author        = {D. Andreuccetti and R. Fossi and C. Petrucci},
  title         = {An Internet resource for the calculation of the dielectric properties of body tissues in the frequency range 10 Hz - 100 GHz},
  year          = {1997},
  note          = {Based on data published by C.Gabriel et al. in 1996.},
  school        = {IFAC-CNR, Florence - Italy},
  urlaccessdate = {01.02.2019},
}

@PhdThesis{Horesh,
  author    = {L Horesh},
  title     = {Some Novel Approaches in Modelling and Image Reconstruction for Multi-Frequency Electrical Impedance Tomography of the Human Brain},
  school    = {Department of Medical Physics - University College London},
  year      = {2006},
}

@Article{Wagshul2011,
  author    = {Mark E Wagshul and Per K Eide and Joseph R Madsen},
  title     = {The pulsating brain: A review of experimental and clinical studies of intracranial pulsatility},
  journal   = {Fluids and Barriers of the {CNS}},
  year      = {2011},
  volume    = {8},
  number    = {1},
  month     = {jan},
  doi       = {10.1186/2045-8118-8-5},
  publisher = {Springer Science and Business Media {LLC}},
}

@Article{JUNJI_2006,
  author    = {Junji Seki and Yasuhiko Satomura and Yasuhiro Ooi and Toshio Yanagidaand Akitoshi Seiyama},
  title     = {Velocity profiles in the rat cerebral microvessels measured by optical coherence tomography},
  journal   = {Clinical Hemorheology and Microcirculation},
  year      = {2006},
  volume    = {34},
  number    = {1-2},
  pages     = {233--9},
  publisher = {IOS Press},
}

@Article{Greitz1992,
  author    = {D. Greitz and R. Wirestam and A. Franck and B. Nordell and C. Thomsen and F. St\aa{}hlberg},
  title     = {Pulsatile brain movement and associated hydrodynamics studied by magnetic resonance phase imaging},
  journal   = {Neuroradiology},
  year      = {1992},
  volume    = {34},
  number    = {5},
  pages     = {370--80},
  doi       = {10.1007/bf00596493},
  publisher = {Springer Science and Business Media {LLC}},
}

@Article{CastelarWembers:759058,
  author    = {Wembers, Carlos C. and Flürenbrock, Fabian and Korn, Leonie Christine and Benninghaus, Anne and Misgeld, Berno Johannes Engelbert and Shchukin, Sergey I. and Radermacher, Klaus and Leonhardt, Steffen},
  title     = {{FEM} {S}imulation of {B}ioimpedance-{B}ased {M}onitoring of {V}entricular {D}ilation and {I}ntracranial {P}ulsation},
  journal   = {International journal of bioelectromagnetism : IJBEM},
  year      = {2019},
  volume    = {21},
  number    = {1},
  pages     = {7--20},
  issn      = {1456-7857},
  address   = {Tampere},
  cid       = {$I:(DE-82)611010_20140620$ / $I:(DE-82)419410_20140620$},
  cin       = {611010 / 419410},
  ddc       = {570},
  publisher = {International Society for Bioelectromagnetism},
  reportid  = {RWTH-2019-03338},
  typ       = {PUB:(DE-HGF)16},
}

@Article{Kneihsl2020,
  author    = {Markus Kneihsl and Edith Hofer and Christian Enzinger and Kurt Niederkorn and Susanna Horner and Daniela Pinter and Simon Fandler-H\"{o}fler and Sebastian Eppinger and Melanie Haidegger and Reinhold Schmidt and Thomas Gattringer},
  title     = {Intracranial Pulsatility in Relation to Severity and Progression of Cerebral White Matter Hyperintensities},
  journal   = {Stroke},
  year      = {2020},
  volume    = {51},
  number    = {11},
  pages     = {3302--9},
  month     = {nov},
  doi       = {10.1161/strokeaha.120.030478},
  publisher = {Ovid Technologies (Wolters Kluwer Health)},
}

@Article{Alastruey2007,
  author    = {J Alastruey and K H Parker and {J et al} Peir{\'{o}}},
  title     = {Modelling the circle of Willis to assess the effects of anatomical variations and occlusions on cerebral flows.},
  journal   = {Journal of Biomechanics},
  year      = {2007},
  volume    = {40},
  number    = {8},
  pages     = {1794--805},
  month     = {jan},
  doi       = {10.1016/j.jbiomech.2006.07.008},
  publisher = {Elsevier {BV}},
}

@article{Dodo2020,
  doi = {10.1016/j.jns.2020.116818},
  year = {2020},
  month = jul,
  publisher = {Elsevier {BV}},
  volume = {414},
  pages = {116818},
  author = {Yoriko Dodo and Tetsuya Takahashi and Kie Honjo and Naoyuki Kitamura and Hirofumi Maruyama},
  title = {Measurement of the length of vertebrobasilar arteries: A three-dimensional approach},
  journal = {Journal of the Neurological Sciences}
}

@article{Fomkina2016,
  doi = {10.15275/rusomj.2016.0205},
  year = {2016},
  month = jun,
  publisher = {{LLC} Science and Innovations},
  volume = {5},
  number = {2},
  pages = {e0205},
  author = {Olga A. Fomkina and Vladimir N. Nikolenko and Elena V. Chernyshkova},
  title = {Morphology and biomechanical properties of cerebellar arteries in adults},
  journal = {Russian Open Medical Journal}
}

@article{Schmitter2013,
  doi = {10.1186/1532-429x-15-s1-w21},
  year = {2013},
  month = jan,
  publisher = {Springer Science and Business Media {LLC}},
  volume = {15},
  number = {S1},
  author = {S Schmitter and BD Jagadeesan and AW Grande and J Sein and K Ugurbil and P V Moortele},
  title = {4D flow measurements in the superior cerebellar artery at 7 Tesla: feasibility and potential for applications in patients with trigeminal neuralgia},
  journal = {Journal of Cardiovascular Magnetic Resonance}
}

@Article{Melis2017,
  author    = {Alessandro Melis and Richard H. Clayton and Alberto Marzo},
  title     = {Bayesian sensitivity analysis of a 1D vascular model with Gaussian process emulators},
  journal   = {International Journal for Numerical Methods in Biomedical Engineering},
  year      = {2017},
  volume    = {33},
  number    = {12},
  pages     = {e2882},
  month     = {may},
  doi       = {10.1002/cnm.2882},
  publisher = {Wiley},
}

@Misc{openBF.jl-2018,
  author       = {Alessandro Melis},
  title        = {openBF: Julia software for 1D blood flow modelling},
  month        = {Oct},
  year         = {2018},
  abstractnote = { openBF is an open-source 1D blood flow solver based on MUSCL finite-volume numerical scheme, written in Julia and released under Apache 2.0 free software license. See https://github.com/INSIGNEO/openBF for the git repository and https://insigneo.github.io/openBF/ for the documentation. },
  doi          = {10.15131/shef.data.7166183},
}

@PhdThesis{MelisThesis,
  author    = {Alessandro Melis},
  title     = {Gaussian process emulators for 1D vascular models},
  month     = {August},
  year      = {2017},
  publisher = {University of Sheffield},
  school    = {University of Sheffield},
}

@Article{Visser1989,
  author    = {K R Visser},
  title     = {Electric properties of flowing blood and impedance cardiography.},
  journal   = {Annals of Biomedical Engineering},
  year      = {1989},
  volume    = {17},
  number    = {5},
  pages     = {463--73},
  month     = {sep},
  doi       = {10.1007/bf02368066},
  publisher = {Springer Science and Business Media {LLC}},
}

@Article{Visser1992,
  author    = {K. R. Visser},
  title     = {Electric conductivity of stationary and flowing human blood at low frequencies},
  journal   = {Medical {\&} Biological Engineering {\&} Computing},
  year      = {1992},
  volume    = {30},
  number    = {6},
  pages     = {636--40},
  month     = {nov},
  doi       = {10.1007/bf02446796},
  publisher = {Springer Science and Business Media {LLC}},
}

@Article{Hoetink,
  author    = {A. E. {Hoetink} and T. J. C. {Faes} and K. R. {Visser} and R. M. {Heethaar}},
  title     = {On the flow dependency of the electrical conductivity of blood},
  journal   = {IEEE Transactions on Biomedical Engineering},
  year      = {2004},
  volume    = {51},
  number    = {7},
  pages     = {1251--61},
  doi       = {10.1109/TBME.2004.827263},
}

@InProceedings{Raaijmakers,
  author    = {E. {Raaijmakers} and J. T. {Marcus} and H. G. Goovaerts and P. M. J. M. Vries and T. J. C. {Faes} and R. M. {Heethaar}},
  title     = {The influence of pulsatile flow on blood resistivity in impedance cardiography},
  booktitle = {Proceedings of 18th Annual International Conference of the IEEE Engineering in Medicine and Biology Society},
  year      = {1996},
  volume    = {5},
  pages     = {1957--8},
  doi       = {10.1109/IEMBS.1996.646338},
}

@Article{Shen2016,
  author    = {Hua Shen and Yong Zhu and Kai-Rong Qin},
  title     = {A theoretical computerized study for the electrical conductivity of arterial pulsatile blood flow by an elastic tube model},
  journal   = {Medical Engineering {\&} Physics},
  year      = {2016},
  volume    = {38},
  number    = {12},
  pages     = {1439--48},
  month     = {dec},
  doi       = {10.1016/j.medengphy.2016.09.013},
  publisher = {Elsevier {BV}},
}

@Article{Shen2018,
  author    = {Hua Shen and Siqi Li and Yu Wang and Kai-Rong Qin},
  title     = {Effects of the arterial radius and the center-line velocity on the conductivity and electrical impedance of pulsatile flow in the human common carotid artery},
  journal   = {Medical {\&} Biological Engineering {\&} Computing},
  year      = {2018},
  volume    = {57},
  number    = {2},
  pages     = {441--51},
  doi       = {10.1007/s11517-018-1889-x},
  publisher = {Springer Nature},
}

@Article{Gaw2008,
  author    = {R. L. Gaw and B. H. Cornish and B. J. Thomas},
  title     = {The Electrical Impedance of Pulsatile Blood Flowing Through Rigid Tubes: A Theoretical Investigation},
  journal   = {IEEE Transactions on Biomedical Engineering},
  year      = {2008},
  volume    = {55},
  number    = {2},
  pages     = {721--7},
  issn      = {0018-9294},
  doi       = {10.1109/TBME.2007.903531},
}

@Article{Bodo2018,
  author    = {Michael Bodo and Leslie D. Montgomery and Frederick J. Pearce and Rocco Armonda},
  title     = {Measurement of cerebral blood flow autoregulation with rheoencephalography: a comparative pig study},
  journal   = {Journal of Electrical Bioimpedance},
  year      = {2018},
  volume    = {9},
  number    = {1},
  pages     = {123--32},
  month     = {dec},
  doi       = {10.2478/joeb-2018-0017},
  publisher = {Walter de Gruyter {GmbH}},
}

@Article{Meghdadi2019,
  author    = {Amir H. Meghdadi and Djordje Popovic and Gregory Rupp and Stephanie Smith and Chris Berka and Ajay Verma},
  title     = {Transcranial Impedance Changes during Sleep: A Rheoencephalography Study},
  journal   = {{IEEE} Journal of Translational Engineering in Health and Medicine},
  year      = {2019},
  volume    = {7},
  pages     = {1--7},
  doi       = {10.1109/jtehm.2019.2898193},
  publisher = {Institute of Electrical and Electronics Engineers ({IEEE})},
}

@Article{Impedancecardiography,
  author    = {Donald P. Bernstein},
  title     = {Impedance cardiography: Pulsatile blood flow and the biophysical and electrodynamic basis for the stroke volume equations},
  journal   = {Journal of Electrical Bioimpedance},
  year      = {2010},
  volume    = {1},
  number    = {1},
  pages     = {2--17},
  address   = {Berlin},
  doi       = {https://doi.org/10.5617/jeb.51},
  publisher = {Sciendo},
}

@InProceedings{Zhang2020,
  author    = {H. J. Zhang and H. Shen and Y. J. Li and K. R. Qin},
  title     = {An in Vitro Circulatory Device for Studying Blood Flow Electrical Impedance in Human Common Carotid Arteries},
  booktitle = {2020 IEEE 16th International Conference on Control Automation (ICCA)},
  year      = {2020},
  pages     = {1518--22},
  doi       = {10.1109/ICCA51439.2020.9264317},
}

@InProceedings{Pesti2019,
  author    = {K. Pesti and H. Kõiv and M. Min},
  title     = {Simulation of the Sensitivity Distribution of Four- Electrode Impedance Sensing on Radial Artery},
  booktitle = {2019 IEEE Sensors Applications Symposium (SAS)},
  year      = {2019},
  pages     = {1--6},
  doi       = {10.1109/SAS.2019.8705976},
}

@Article{Badeli2020,
  author    = {V. Badeli and G. M. Melito and A. R. Köstinger and O. B{\'{\i}}r{\'{o}} and K. Ellermann},
  title     = {Electrode positioning to investigate the changes of the thoracic bioimpedance caused by aortic dissection {\textendash} a simulation study},
  journal   = {Journal of Electrical Bioimpedance},
  year      = {2020},
  volume    = {11},
  number    = {1},
  pages     = {38--48},
  month     = {jan},
  doi       = {10.2478/joeb-2020-0007},
  publisher = {Walter de Gruyter {GmbH}},
}

@Article{Braun_2018,
  author    = {Fabian Braun and Martin Proen{\c{c}}a and Mathieu Lemay and Mattia Bertschi and Andy Adler and Jean-Philippe Thiran and Josep Sol{\`{a}}},
  title     = {Limitations and challenges of {EIT}-based monitoring of stroke volume and pulmonary artery pressure},
  journal   = {Physiological Measurement},
  year      = {2018},
  volume    = {39},
  number    = {1},
  pages     = {014003},
  doi       = {10.1088/1361-6579/aa9828},
  publisher = {{IOP} Publishing},
}

@InProceedings{rgb,
  author    = {R. G. Beraldo and F. S. Moura},
  title     = {Time-difference Electrical Impedance Tomography with a Blood Flow Model as Prior Information for Stroke Monitoring},
  booktitle = {Proceedings of the XXVII Brazilian Congress on Biomedical Engineering},
  pages = {280--5},
  year      = {2020},
  address   = {Vitória, Brazil},
  publisher = {SBEB},

}

@Article{Goren2018,
  author    = {Nir Goren and James Avery and Thomas Dowrick and Eleanor Mackle and Anna Witkowska-Wrobel and David Werring and David Holder},
  title     = {Multi-frequency electrical impedance tomography and neuroimaging data in stroke patients},
  journal   = {Scientific Data},
  year      = {2018},
  volume    = {5},
  number    = {1},
  month     = {jul},
  doi       = {10.1038/sdata.2018.112},
  publisher = {Springer Science and Business Media {LLC}},
}

@Article{Vikner2019,
  author    = {Tomas Vikner and Lars Nyberg and Madelene Holmgren and Jan Malm and Anders Eklund and Anders W{\aa}hlin},
  title     = {Characterizing pulsatility in distal cerebral arteries using 4D flow {MRI}},
  journal   = {Journal of Cerebral Blood Flow {\&} Metabolism},
  year      = {2019},
  volume    = {40},
  number    = {12},
  pages     = {2429--40},
  month     = {nov},
  doi       = {10.1177/0271678x19886667},
  publisher = {{SAGE} Publications},
}

@Article{Holmgren2019,
  author    = {Madelene Holmgren and Anders W{\aa}hlin and Tora Dun{\aa}s and Jan Malm and Anders Eklund},
  title     = {Assessment of Cerebral Blood Flow Pulsatility and Cerebral Arterial Compliance With 4D Flow {MRI}},
  journal   = {Journal of Magnetic Resonance Imaging},
  year      = {2019},
  volume    = {51},
  number    = {5},
  pages     = {1516--25},
  month     = {nov},
  doi       = {10.1002/jmri.26978},
  publisher = {Wiley},
}

@Article{Kucewicz2008,
  author    = {John C. Kucewicz and Barbrina Dunmire and Nicholas D. Giardino and Daniel F. Leotta and Marla Paun and Stephen R. Dager and Kirk W. Beach},
  title     = {Tissue Pulsatility Imaging of Cerebral Vasoreactivity During Hyperventilation},
  journal   = {Ultrasound in Medicine {\&} Biology},
  year      = {2008},
  volume    = {34},
  number    = {8},
  pages     = {1200--8},
  month     = {aug},
  doi       = {10.1016/j.ultrasmedbio.2008.01.001},
  publisher = {Elsevier {BV}},
}

@Article{Desmidt2018,
  author    = {Thomas Desmidt and Fr{\'{e}}d{\'{e}}ric Andersson and Bruno Brizard and Paul-Armand Dujardin and Jean-Philippe Cottier and Fr{\'{e}}d{\'{e}}ric Patat and Jean-Pierre R{\'{e}}m{\'{e}}ni{\'{e}}ras and Val{\'{e}}rie Gissot and Wissam El-Hage and Vincent Camus},
  title     = {Ultrasound Measures of Brain Pulsatility Correlate with Subcortical Brain Volumes in Healthy Young Adults},
  journal   = {Ultrasound in Medicine {\&} Biology},
  year      = {2018},
  volume    = {44},
  number    = {11},
  pages     = {2307--13},
  month     = {nov},
  doi       = {10.1016/j.ultrasmedbio.2018.06.016},
  publisher = {Elsevier {BV}},
}

@Article{GMSH2009,
  author    = {Christophe Geuzaine and Jean François Remacle},
  journal   = {International Journal for Numerical Methods in Engineering},
  title     = {Gmsh: A 3-D finite element mesh generator with built-in pre- and post-processing facilities},
  year      = {2009},
  month     = may,
  number    = {11},
  pages     = {1309--31},
  volume    = {79},
  doi       = {10.1002/nme.2579},
  publisher = {Wiley},
}

@Unpublished{turovets,
  author    = {D. Hammond  and N. Price and S. Turovets},
  title     = {Construction and segmentation of pediatric head tissue atlases for electrical head modeling},
  note      = {OHBM, Vancouver, Canada},
  year      = {2017},
}

@Unpublished{turovets2,
  author    = { Jidong Hou and Sergei Turovets  and Kai Li and Phan Luu and Don Tucker and Linda Larson-Prior},
  title     = {Spatially Resolved Pediatric Skull Conductivities for Inhomogeneous Electrical Forward Modeling},
  note      = {OHBM, Vancouver, Canada},
  year      = {2017},
}

@Article{turovets3,
  author    = {Jasmine Song and Kyle Morgan and Sergei Turovets and Kai Li and Colin Davey and Pavel Govyadinov and Phan Luu and  Kirk Smith and Fred Prior and Linda Larson-Prior and Don M. Tucker},
  title     = {Anatomically Accurate Head Models and Their Derivatives for Dense Array EEG Source Localization},
  journal   = {Funct Neurol Rehabil Ergon},
  year      = {2013},
  volume    = {3},
  pages     = {275--93},
  number     = {2-3},
}

@Article{mha,
  author    = {E. Bullitt and Donglin Zeng and Guido Gerig and Stephen Aylward and Sarang Joshi and J. Keith Smith and Weili Lin and Matthew G. Ewend},
  title     = {Vessel tortuosity and brain tumor malignancy: A blinded study},
  journal   = {Academic Radiology},
  year      = {2005},
  volume    = {12},
  pages     = {1232--40},
  doi       = {10.1016/j.acra.2005.05.027},
  numer     = {10},
}

@Article{avants2009ants,
  author    = {Avants, BB and Tustison, NJ and Song, G and Gee, JC},
  title     = {Ants: Open-source tools for normalization and neuroanatomy.},
  journal   = {HeanetIe},
  year      = {2009},
  volume    = {10},
  pages     = {1--11},
}

@Article{AVANTS2008,
  author    = {B Avants and C Epstein M Grossman and J Gee},
  title     = {Symmetric diffeomorphic image registration with cross-correlation: Evaluating automated labeling of elderly and neurodegenerative brain},
  journal   = {Medical Image Analysis},
  year      = {2008},
  volume    = {12},
  number    = {1},
  pages     = {26--41},
  month     = {feb},
  doi       = {10.1016/j.media.2007.06.004},
  publisher = {Elsevier {BV}},
}

@Article{fonov2009unbiased,
  author    = {Fonov, VS and Evans, AC and McKinstry, RC and Almli, CR and Collins, DL},
  title     = {Unbiased nonlinear average age-appropriate brain templates from birth to adulthood.},
  journal   = {NeuroImage},
  year      = {2009},
  volume    = {47},
  number    = {1},
  pages     = {S102},
}

@InProceedings{Grabner_2006,
  author    = {Grabner, G{\"u}nther and Janke, Andrew L. and Budge, Marc M. and Smith, David and Pruessner, Jens and Collins, D. Louis},
  title     = {Symmetric Atlasing and Model Based Segmentation: An Application to the Hippocampus in Older Adults},
  booktitle = {Medical Image Computing and Computer-Assisted Intervention -- MICCAI 2006},
  year      = {2006},
  editor    = {Larsen, Rasmus and Nielsen, Mads and Sporring, Jon},
  pages     = {58--66},
  address   = {Berlin, Heidelberg},
  publisher = {Springer Berlin Heidelberg},
}

@Article{ashburner2005unified,
  author    = {Ashburner, J and Friston, KJ},
  title     = {Unified segmentation.},
  journal   = {Neuroimage},
  year      = {2005},
  volume    = {26},
  number    = {3},
  pages     = {839--51},
  publisher = {Elsevier},
}
\end{document}